\documentclass[twocolumn]{aastex631}

\usepackage{hyperref}
\usepackage{amsmath}
\usepackage{soul}
\usepackage{enumitem}

\begin{document}

\title{SuNeRF-CME: Physics-Informed Neural Radiance Fields for Tomographic Reconstruction of Coronal Mass Ejections}

\author[0000-0002-9309-2981]{Robert Jarolim}
\affiliation{High Altitude Observatory, NSF National Center for Atmospheric Research, USA}
\affiliation{Institute of Physics, University of Graz, Austria}

\author{Martin Sanner}
\affiliation{School of Science and Engineering, University of Dundee, UK}

\author[0000-0001-5823-5783]{Chia-Man Hung}
\affiliation{Oxford Robotics Institute, University of Oxford, UK}

\author{Emma Stevenson}
\affiliation{Applied Intelligence and Data Analysis Group, School of Computer Science, Universidad Politécnica de Madrid, Spain}

\author[0000-0002-8091-0367]{Hala Lamdouar}
\affiliation{Institute of Biomedical Engineering, Department of Engineering Science, University of Oxford, UK}

\author[0000-0003-2780-7843]{Josh Veitch-Michaelis}
\affiliation{Department of Information Technology and Electrical Engineering, ETH Zürich, Switzerland}

\author[0009-0002-6807-2951]{Ioanna Bouri}
\affiliation{Department of Computer Science, University of Helsinki, Finland}

\author[0000-0001-5066-8509]{Anna Malanushenko}
\affiliation{High Altitude Observatory, NSF National Center for Atmospheric Research, USA}

\author[0000-0001-8875-7478]{Elena Provornikova}
\affiliation{Applied Physics Laboratory, Johns Hopkins University, USA}

\author[0000-0001-6558-7197]{Vít Růžička}
\affiliation{Department of Computer Science, University of Oxford, UK}

\author{Carlos Urbina-Ortega}
\affiliation{European Space Research and Technology Centre (ESTEC), European Space Agency, Netherlands}



\begin{abstract}

Coronagraphic observations enable direct monitoring of coronal mass ejections (CMEs) through scattered light from free electrons, but determining the 3D plasma distribution from 2D imaging data is challenging due to the optically-thin plasma and the complex image formation.
We introduce SuNeRF-CME, a framework for 3D tomographic reconstructions of the heliosphere using multi-viewpoint coronagraphic observations. The method leverages Neural Radiance Fields (NeRFs) to estimate the electron density in the heliosphere through ray tracing, while accounting for the underlying Thomson scattering. The model is optimized by iteratively fitting the time-dependent observational data. In addition, we apply physical constraints in terms of continuity, propagation direction, and speed of the heliospheric plasma to overcome limitations imposed by the sparse number of viewpoints.
We utilize synthetic observations of a CME simulation to quantify the model's performance for different viewpoint configurations. Within this controlled synthetic setting, the results demonstrate that our method can reliably estimate the CME parameters from only two viewpoints, with a mean velocity error of $3.01\pm1.94\%$ and propagation direction errors of $3.39\pm1.94^\circ$ in latitude and $1.76\pm0.79^\circ$ in longitude. We further show that our approach can achieve a 3D reconstruction of the simulated CME from two viewpoints, where we correctly model the three-part structure, deformed CME front, and internal plasma variations.
Additional viewpoints can be seamlessly integrated, directly enhancing the reconstruction of the plasma distribution in the heliosphere. These results demonstrate the potential of physics-informed radiance-field methods for CME tomography, paving the way for future extensions toward observational data and space weather applications.

\end{abstract}



\section{Introduction}
\label{sec:intro}

In the upper solar atmosphere, the solar corona, emerging magnetic flux and shearing motions can store large amounts of magnetic energy, which gradually build up over hours to days, and can result in the sudden release within minutes \citep{forbes2006cme, Wiegelmann2014}. This energy release, in the form of solar flares \citep{fletcher2011}, is frequently associated with the ejection of coronal plasma, Coronal Mass Ejections \citep[CMEs; ][]{webb2012cmes}. During this process, coronal plasma is jointly ejected with the magnetic field into interplanetary space, transitioning from a magnetically dominated regime, to a plasma (solar wind) dominated regime. In their early evolution, CMEs can undergo deformations, deflections, or result in failed eruptions \citep{temmer2023propagation}. During the propagation through interplanetary space, CMEs further interact with the ambient plasma, leading to further changes in propagation direction, velocity, and potential deformation \citep[e.g., solar wind drag; ][]{cargill2004drag, vrsnak2013drag}. CMEs also pose a significant space weather risk \citep{temmer2021space_weather, Pulkkinen2007space_weather}, where the precise estimation of speeds, propagation direction, and total mass, are essential to provide inputs for heliospheric models \citep{pomoell2018euhforia, odstrcil2003enlil, moestl2015elevo, amerstorfer2021elevohi, bauer2021elevohi} and space-weather forecasts.

To investigate the processes driving CME propagation and to assess their potential space-weather impacts, it is essential to obtain a 3D characterization of the plasma density in the solar corona and interplanetary space. Direct interpretation of observational data, however, is typically hindered by significant limitations.
Observations of propagating CMEs are primarily obtained from sparse in-situ measurements, detected radio signatures, and remote sensing with white-light coronagraphs aboard space- and ground-based facilities \citep[see][and references therein]{webb2012cmes}. Coronagraphs employ an occulting disk to block the bright solar disk and reveal the faint emission in the solar corona. These observations originate from Thomson scattered light of free electrons in the solar corona \citep{billings1966guide}, and enable a direct monitoring of electron density and its dynamic evolution in the heliosphere (e.g., CMEs). However, interpreting this data is inherently challenging due to the complexity of the image formation process. The observed radiances (surface brightnesses) are line-of-sight integrals through optically thin plasma, and the Thomson scattering process has a complex viewpoint dependence, which prevents a straightforward interpretation of the imaging data. In particular, the projection of the 3D structures to the 2D plane-of-sky can lead to an underestimation of the speed and overestimation of the CME angular width \citep{Burkepile2004projection, vrsnak2007projection, temmer2009projection}.

Various methods aim to estimate CME characteristics through manual fitting \citep{Thernisien2011ApJSgcs, temmer2009projection}, usage of polarized observations \citep{deKonig2009cme_localization, deKonig2011polarimetric_localization, susino2016cme_parameters, deForest2017chirality} or by applying stereoscopic assumptions \citep{Aschwanden2011stereoscopy}. The Graduated Cylindrical Shell (GCS) reconstruction method is a widely applied and practical approach where the shape of the CME is approximated as a cone and the white-light observations are used to find the best matching orientation of the 3D cone. This allows for the estimation of the CME orientation and propagation direction directly from the imaging data \citep{Thernisien2006gcs, Thernisien2011ApJSgcs}. While the GCS approach has proven effective for many events, it has inherent limitations when applied to more complex CME shapes (e.g., deformed front), and the interpretation of the observations can introduce systematic deviations. Specifically, the line-of-sight integrated scattered light is treated as a proxy for the actual plasma distribution, which limits applicability and introduces deviations from the ground-truth distribution. Improvements can be achieved by using observations from multiple viewpoints and jointly fitting the 3D model from two perspectives. \cite{2023AdSpR..72.5243V} showed that for an idealized test set, the average GCS fitting error for two viewpoints with a 60$^\circ$ separation angle is in the range of 7.8$^\circ$ in latitude and 6.8$^\circ$ in longitude. Similarly stereoscopic and triangulation methods aim to directly reconstruct the CME front using observations from multiple viewpoints \citep{Aschwanden2011stereoscopy}. However, for Thomson scattered light, there is a significant dependence on the geometry which needs to be accounted for \citep{Vourlidas2006proper_treatment}. A more consistent approach is the use of observations of polarized and total brightness, where the brightness ratio can be used to determine the center of mass along the line of sight in a physically consistent way \citep{Colaninno2009mass_determination, deKonig2011polarimetric_localization}. While this approach can provide good estimates for dense plasma regions, in practice CMEs are extended 3D structures and the background solar wind plays a significant role \citep{howard2015cme_detectability}.

Tomographic methods address these limitations by inverting multiple 2D observations to recover the 3D plasma distribution while accounting for Thomson scattering. With a single viewpoint, the inversion problem is inherently ill-posed, and classical coronal tomography therefore exploits solar rotation to provide additional constraints for structures that evolve slowly in time \citep{frazin2005tomography}. For dynamic phenomena, static assumptions lead to temporal smearing and reconstruction artifacts. Consequently, time-dependent extensions have been developed using temporal regularization \citep{vibert2016tomography}, state-estimation techniques such as Kalman filtering \citep{frazin2005time_tomography, butala_2008dynamic_tomography, butala2010dynamic_tomography}, and alternative basis representations such as spherical harmonics \citep{morgan2019tomography_spherical_harmonics}.

\citet{ramos2023tomography} demonstrated that neural representations can perform time-dependent tomographic reconstructions of the solar corona with favorable implicit regularization for spatiotemporal coherence and high parameter efficiency. For CME-specific applications, \citet{frazin2009cme_tomography} reconstructed a two-dimensional slice of a simulated CME using three viewpoints and level-set techniques. \citet{jackson2006smei_tomography} developed an iterative tomographic framework that combines interplanetary scintillation (IPS) velocity data with white-light observations and kinematic constraints, enabling reconstructions of CME morphology and propagation together with the background solar wind \citep{jackson2011tomography_review}.

Even after accounting for all underlying physical processes, ghost-trajectories can lead to ambiguous results \citep{deforest2013ghost}. Therefore, CME parameter estimates are typically associated with large uncertainties, and interactions in interplanetary space are still incompletely characterized.

In this study, we present a novel deep learning approach for tomographic reconstructions of CMEs based on multi-viewpoint data. For this we build on our previous approach, which leveraged Neural Radiance Fields \citep[NeRFs;][]{mildenhall2021nerf} for tomographic reconstructions of the EUV corona \citep[Sun Neural Radiance Fields; SuNeRF;][]{jarolim2024sunerf}, and adapt it for a physically consistent treatment of white-light coronagraphic observations (SuNeRF-CME). In this work, the neural network is used as a continuous implicit parameterization of a tomographic inverse problem, rather than as a generative or predictive model trained on a population of samples. The neural network serves as a function representation of the reconstruction volume, mapping coordinate points to their respective physical quantities. The network parameters are optimized to reproduce a fixed set of white-light observations through a known physical forward model.
We further leverage Physics-Informed Neural Networks \citep[PINNs; ][]{raissi2019pinns} which allow us to incorporate additional physical constraints to mitigate challenges due to the limited number of observations. With this, we provide a first implementation of a Physics-Informed Neural Radiance Field (Sect. \ref{sec:method_data}). The primary aim of this study is to quantify how reconstruction performance depends on the available observational information and on the inclusion of physics-informed constraints. We utilize synthetic white-light observations of a CME to validate our approach. The synthetic observations enable the analysis of various viewpoint configurations and provide a quantification of the model performance by comparing the ground-truth plasma distribution with our reconstruction. Note that this study is limited to synthetic data only, while we leave the extension to observational data for a future study.

\section{Method}
\label{sec:method}

\begin{figure}
    \centering
    \includegraphics[width=\linewidth]{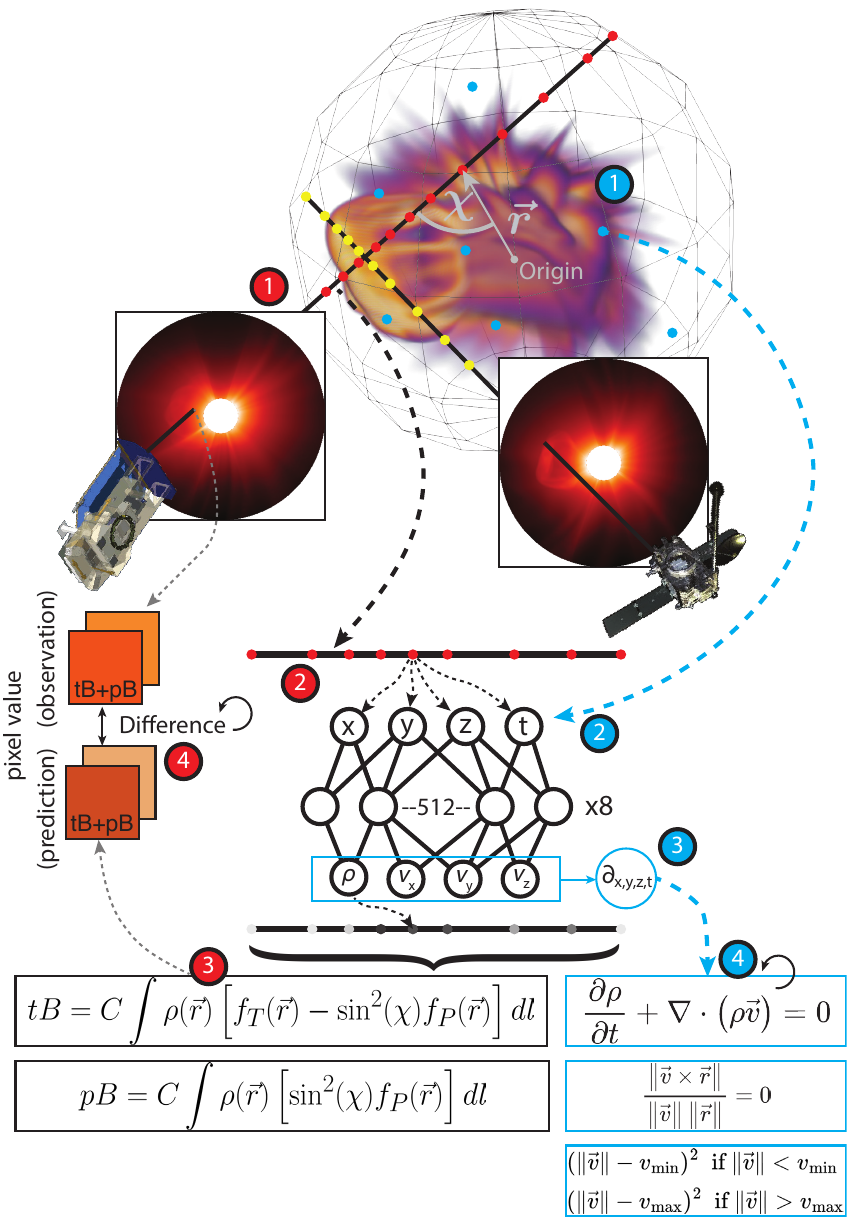}
    \caption{Overview of the SuNeRF-CME method for tomographic reconstruction of coronal mass ejections (CMEs) using physics-informed Neural Radiance Fields. The approach leverages multi-viewpoint observations of total $tB$ and polarized $pB$ brightness to estimate the electron density in the heliosphere. The tomographic reconstruction is performed via a ray tracing approach (red):
    (1) Rays are traced per pixel into the reconstruction volume, sampling spatial-temporal coordinates $x, y, z, t$ along the line of sight.
    (2) A neural network maps these coordinates to local electron density $\rho$ and plasma velocity $\vec{v} = (v_x, v_y, v_z)$.
    (3) The electron density and coordinate positions are used to compute Thomson-scattered radiance, integrating along the line of sight.
    (4) The resulting estimated total and polarized brightness values are compared to the observations, and the model is updated to minimize the deviation.
    To address challenges posed by sparse viewpoint coverage, additional physical constraints are incorporated (blue):
    (1) Points are randomly sampled within the 3D domain.
    (2) The neural network maps the coordinate points to electron density and velocity.
    (3) Using automatic differentiation, the derivatives of the output values with respect to the input coordinates are constructed.
    (4) In combination with the ray tracing loss, the model is constrained by minimizing the residuals of the continuity equation, regularizing for an approximately radial solar wind, and enforcing consistency with an expected ambient solar wind speed range.
    }
    \label{fig:method}
\end{figure}

For our tomographic reconstruction method we follow \cite{billings1966guide} to obtain the total and polarized brightness observations from the modeled plasma density (Sect. \ref{sec:tomography}). We build on NeRFs to perform the ray tracing approach (Sect. \ref{sec:nerf}) and apply additional physics constraints through a PINN (Sect. \ref{sec:pinn}). We use the combined approach (Sect. \ref{sec:sunerf}) to perform physics-informed tomographic reconstructions that combine the white-light observations with basic assumptions about the solar wind propagation (i.e., continuity equation). In this study, we focus solely on synthetic CME observations (Sect. \ref{sec:method_data}), which are used to quantify the performance of our approach by comparing to the ground-truth plasma distribution (Sect. \ref{sec:evaluation}).

\subsection{Tomography}
\label{sec:tomography}

White-light coronagraph observations are obtained from  the Thomson scattered light of free electrons in the heliosphere \citep{howard2009thomson_theory}. The detected signal does not originate from local plasma emission, but from the elastic scattering of photons in the low-energy regime, and therefore depends on the geometric configuration between the  scattering location (i.e., the free electron), the source of the radiation (i.e., the solar photosphere), and the observer. The observed quantity corresponds to the radiance (surface brightness) measured per image pixel, corresponding to the line-of-sight integral of the Thomson-scattered signal through the optically thin plasma. Following \cite{2012ApJ...752..130H}, radiance denotes the quantity measured by an imaging detector per solid angle element, whereas intensity refers to the radiance integrated over the apparent area of a feature. In this work, all forward-modeled and observed image quantities correspond to radiance.
We adopt the geometrical formulation of \citet{billings1966guide}. In this framework, the observed radiance depends on the spatial location of the scattering point $\vec{r}$, the scattering angle $\chi$ (i.e., the angle between the incident solar radiation and the line of sight), and the electron density $\rho(\vec{r})$. The total $tB$ and polarized $pB$ brightness observed are given by the integral of the Thomson-scattered radiance along the line of sight $l$

\begin{equation}
    \label{eq:pB}
    pB = C \int \rho(\vec{r}) \left[\sin^2(\chi) f_P(\vec{r}) \right] dl \,,
\end{equation}

\begin{equation}
    \label{eq:tB}
    tB = C \int \rho(\vec{r}) \left[ f_T(\vec{r}) - \sin^2(\chi) f_P(\vec{r}) \right] dl \,.
\end{equation}
Here, $tB$ corresponds to the Stokes parameter $I$, while $pB = B_T - B_R$ represents a Stokes $Q$ parameter evaluated in a coordinate system rotated into the local radial--tangential scattering frame, where $B_T$ and $B_R$ denote radiances measured through tangential and radial polarizers, respectively \citep[cf.][]{2022ApJ...927...98D, Gibson2025PolarizationDiagnostics}.
$C$ is a constant scaling factor given by the electron cross section $\sigma_e=7.95\times10^{-26}$ cm$^2$ sr$^{-1}$, and the solar brightness $I_0=2.3\times10^{11}$ W cm$^{-2}$ sr$^{-1}$,
\begin{equation}
    C = I_0\frac{\pi \sigma_e}{2} \,.
\end{equation}
$\chi$ is the scattering angle $\sin^2(\chi)=\lVert \hat{l} \times \hat{r}\rVert^2$. The functions $f_T(\vec{r})$ and $f_P(\vec{r})$ describe the radiance variation based on the spatial location to the scattering point $\vec{r}$
\begin{equation}
    f_P(\vec{r}) = (1 - u) A + u B \,,
\end{equation}
\begin{equation}
    f_T(\vec{r}) = 2 [(1 - u)C + uD] \,.
\end{equation}
Here, $u$ refers to the limb-darkening coefficient with $u\approx0.63$, and $A$, $B$, $C$, and $D$ are functions of the angle between the distance of the scattering point from the solar center $r=\left\lVert \vec{r} \right\rVert$ and the solar radius $R_\odot$
\begin{equation}
    \Omega=\arcsin(R_\odot/r) \,,
\end{equation}

\begin{equation}
A = \cos \Omega \sin^2 \Omega,
\end{equation}

\begin{equation}
\begin{split}
B = -\frac{1}{8} \Bigg[ 1 - 3 \sin^2 \Omega
    - \frac{\cos^2 \Omega}{\sin \Omega} (1 + 3 \sin^2 \Omega) \\
    \times \ln \left( \frac{1 + \sin \Omega}{\cos \Omega} \right) \Bigg],
\end{split}
\end{equation}

\begin{equation}
C = \frac{4}{3} - \cos \Omega - \frac{\cos^3 \Omega}{3},
\end{equation}

\begin{equation}
\begin{split}
D = \frac{1}{8} \Bigg[ 5 + \sin^2 \Omega
    - \frac{\cos^2 \Omega}{\sin \Omega} (5 - \sin^2 \Omega) \\
    \times \ln \left( \frac{1 + \sin \Omega}{\cos \Omega} \right) \Bigg].
\end{split}
\end{equation}



The coefficients $A$, $B$, $C$, and $D$ are the classical van de Hulst coefficients \citep{vandeHulst1950}, which account for the finite angular extent of the solar disk and limb darkening of the solar radiation. In the small solid-angle limit (i.e., at sufficiently large heliocentric distances, $\Omega \ll 1$), $A$, $C$, and $D$ vanish, and the formulation reduces to the point-source approximation, which is typically valid beyond $\sim 2$--$5\,R_\odot$ depending on the required accuracy \citep{2012ApJ...752..130H}. For our implementation, we retain the full set of van de Hulst coefficients, such that the point-source limit is recovered naturally at large heliocentric distances, while close to the Sun these effects remain non-negligible.

For spatially extended plasma such as the background solar wind, white-light polarization primarily diagnoses the radial density falloff rather than a unique three-dimensional position along the line of sight \citep{Gibson2025PolarizationDiagnostics}, provided that the scattering angle $\chi$ varies substantially along the line of sight (i.e., by a significant fraction of $\pi$ within the integrals in Equations \ref{eq:pB} and \ref{eq:tB}). In contrast, sharp density enhancements at CME leading edges and internal substructures fall within the localized regime, for which polarization provides direct geometric constraints.


Tomography provides a tool that can consistently combine multi-viewpoint data into a 3D representation of the heliosphere \citep[e.g.,][]{frazin2005tomography, Aschwanden2011stereoscopy}. By using a physically consistent forward model, this approach can account for the underlying physics of the image formation, providing reconstructions of both CMEs and the overall plasma distribution in the heliosphere. A fundamental challenge of tomographic inversion is the large number of degrees of freedom, which can lead to under-constrained or non-unique solutions when only a limited number of simultaneous observations are available \citep{Burkepile2004projection}. Consequently, classical tomography has been primarily applied to quasi-static structures in the inner corona, where solar rotation provides additional viewing angles over time \citep{frazin2005tomography, vasquez2008srt}. Extension to time-dependent reconstructions requires additional constraints, either through explicit regularization \citep{frazin2005time_tomography, frazin2009cme_tomography, jackson2011tomography_review} or implicit smoothness enforced by the model representation \citep{ramos2023tomography} or a prescribed set of basis functions \citep{morgan2019tomography_spherical_harmonics}. In this study, we combine the implicit spatiotemporal smoothness of neural representations with explicit physics-based constraints (Sect.~\ref{sec:sunerf}) to mitigate the ill-posed nature of the inverse problem.


\subsection{Neural Radiance Fields}
\label{sec:nerf}

NeRFs are a ray tracing approach, where a 3D scene is reconstructed from a set of images and their known observer positions. A key advantage of a NeRF is that the reconstructed volume is represented by a Neural Network, which maps coordinate points to the respective physical quantities, instead of using an explicit grid-representation. Radiance-field approaches such as voxel-based representations \citep[e.g., Plenoxels][]{fridovich2022plenoxels} and point-based methods \citep[e.g., Gaussian splatting][]{kerbl20233d} provide alternatives to NeRFs; however, they are less naturally compatible with the continuous differential constraints employed here (Sect.~\ref{sec:pinn}). For a classical NeRF approach, the modeled quantities are given by the opacity and color. After fitting the model, novel viewpoints can be rendered from the reconstructed 3D scene. Note that in contrast to other deep learning approaches, such as large pretrained models or methods trained on diverse datasets across many scenes, this method does not rely on a pre-existing dataset. Instead, it reconstructs a specific scene directly from the available multi-vantage observational data, focusing on reconstructing that scene rather than generalizing across multiple scenes. Therefore, to obtain a 3D reconstruction from a novel set of observations, a new model needs to be trained from scratch.

 However, solar rotation-based tomography relies on the assumption of quasi-static structures, which is violated for rapidly evolving phenomena such as CMEs and leads to temporal smearing and reconstruction artifacts.

In solar physics NeRFs were applied to reconstruct the emission and absorption profile of the solar corona by using multi-vantage point observations and the solar rotation \citep{jarolim2024sunerf}. Directly related to our study, \cite{ramos2023tomography} used a NeRF approach in the context of solar rotational tomography to estimate the solar K-corona from single-viewpoint observations. While such approaches enable time-dependent reconstructions, they remain limited by their reliance on solar rotation, where structural evolution during the observation window can introduce temporal smearing and reconstruction ambiguities for rapidly evolving structures.


\subsection{Physics-Informed Neural Networks}
\label{sec:pinn}

Physics-Informed Neural Networks (PINNs) are conceptually related to NeRFs in that both use neural networks to represent a continuous volume by mapping spatial coordinates to physical quantities. However, the core distinction lies in how they are trained and applied. NeRFs are typically optimized to overfit a small number of multi-vantage-point observations, allowing them to reconstruct a specific 3D scene by interpolating between the views. In contrast, PINNs are trained to solve boundary value problems, where the governing set of partial differential equations (PDEs) is used to extrapolate the sparse observational data. Neural networks are fully differentiable functions, therefore the neural representation can be used directly to compute smooth derivatives of the output quantities with respect to the input coordinates. By minimizing the residuals of the PDEs and satisfying boundary conditions, PINNs provide smooth and physically constrained solutions that can incorporate noisy or incomplete data. This makes them particularly useful for data-driven simulations where observational data alone are insufficient.

In solar physics, PINNs have been applied to magnetic field modeling \citep{jarolim2023nf2, Jarolim2024multi}, radiative hydrodynamic simulations \citep{keller2025hd_PINN}, and spectropolarimetric inversion \citep{base2025inversion, jarolim2025pinnme}. 
In this work, we use PINNs to integrate physical constraints into the NeRF framework, thereby overcoming its inherent limitation of poor extrapolation for a limited number of viewpoints.

\subsection{Physics-Informed Neural Radiance Field}
\label{sec:sunerf}
In this study, we combine the concepts of NeRFs and PINNs for the tomographic reconstruction of CMEs. Our approach leverages the concept of a neural representation, which enables a memory-efficient representation of the 4D volume. Specifically, the neural representation can encode the full temporal evolution of the 3D volume in $\sim$1.8 million free parameters. The function representation of the neural network implies an intrinsic spatiotemporal smoothness, which mitigates spontaneous fluctuations in the reconstructed plasma distribution \citep{ramos2023tomography, jarolim2024sunerf, jarolim2025pinnme}. We adopt the full spatially dependent formulation of Thomson scattering to account for variations in scattering efficiency across the line of sight. In addition, we incorporate physics-informed constraints (PINN) to mitigate unphysical reconstructions due to limited viewpoints by enforcing physical consistency across the time dimension.

\subsubsection{Model}
Given that coronal tomography is generally poorly constrained by limited observations, we surmised that the model architecture (e.g., number of layers, neurons) is of secondary importance, and the quality of the reconstruction primarily depends on the available input data. Therefore, we select a standard SIREN architecture, consisting of eight hidden layers with 512 neurons each. This configuration ought to provide a sufficient number of trainable parameters to represent the 3D simulation volume, consistent with related NeRF implementations \citep{mildenhall2021nerf, jarolim2024sunerf}. We use the default SIREN configuration with $w_0=30$ for the first layer and $w_0=1$ for the remaining layers  \citep[c.f.,][]{sitzmann2020siren}. 

\subsubsection{Ray Tracing}
We build on our previous tomography method for EUV observations \citep[SuNeRF; ][]{jarolim2024sunerf}, which uses a NeRF approach to estimate the 3D emission and absorption profile in the solar corona. Here, we implement the tomographic reconstruction of white-light coronagraphic observations of polarized and total brightness based on Eq. \ref{eq:pB} and \ref{eq:tB}. Figure \ref{fig:method} provides an overview of our method. For each observation we cast rays through each image pixel, where we use the helioprojective angles and their known orbital position $\vec{r}_\text{Obs}=(r, \theta, \phi)$, where $r$ corresponds to the radius, $\theta$ to the latitude, and $\phi$ to the longitude in Heliocentric coordinates. Note that this contrasts with the employed Carrington frame in \cite{jarolim2024sunerf}. The ray direction is computed such that the central coordinate points are positioned at the image center.
For each ray we sample points in a spherical shell of 130 solar radii around the Sun. Each coordinate point ($x, y, z, t$) is mapped through the neural network to the corresponding logarithmic electron density $\ln{\rho}$, and the velocity vector $\vec{v}=(v_x, v_y, v_z)$. Defining the neural network output as the logarithm of the plasma density ensures physically meaningful (non-negative) densities throughout the reconstruction domain.

The output electron density is scaled by a radial drop-off in accordance to an expected solar wind profile $\ln{\rho} =\ln{\rho}’ – 2 \ln{r}$, where $r$ corresponds to the radial distance of the sampling point from the Sun center, and $\ln{\rho}’$ denotes the model-predicted logarithmic density. The final electron density is then obtained by $\rho=e^{\ln{\rho}}$. The velocity is modeled as a perturbation to a base radial solar-wind profile
\begin{equation}
\vec{v} = \vec{v}_0 + \vec{v}’,
\end{equation}
where $\vec{v}_0 = 300~\mathrm{km\,s^{-1}}\,\hat{r}$ represents the background radial solar-wind speed, and $\vec{v}' = (v_x',\,v_y',\,v_z')$ denotes the model-predicted velocity perturbation.

\subsubsection{Forward Rendering}
Based on the modeled electron density we compute the total and polarized brightness according to Eq. \ref{eq:pB} and \ref{eq:tB}, where $C=100$ is selected to provide a computationally efficient range. Therefore, the calculations are performed in model units for the density, and the resulting quantities are rescaled to physical units for evaluation. For the integration, we use finite differences, where $dl$ corresponds to the spatial distance between adjacent sampling points.

The pixel brightness (radiance) values are computed in two steps, where we first sample 64 points across the full ray path with an equal spacing and random perturbations (stratified sampling). The resulting tB and pB brightness values are then used to sample 128 points weighted by the radiance distribution along the ray (hierarchical sampling). Both samplings are performed with the same NeRF model, in contrast to a coarse and fine model \citep[c.f.][]{jarolim2024sunerf, mildenhall2021nerf}. This strategy provides comparable performance to separate coarse/fine models while reducing computational requirements. The hierarchical sampling strategy enables a higher sampling in regions that strongly contribute to the observed emission, resulting in a better resolved reconstruction \citep{mildenhall2021nerf}.

\subsubsection{Image-based Loss Functions}
For the image-based losses, the resulting total and polarized brightness values are compared to the observations. We minimize the difference between the logarithmically scaled total and polarized brightness,
\begin{equation}
L_{\mathrm{tB}} = (\ln I_{\mathrm{tB,obs}} - \ln I_{\mathrm{tB,pred}})^2,
\end{equation}
\begin{equation}
L_{\mathrm{pB}} = (\ln I_{\mathrm{pB,obs}} - \ln I_{\mathrm{pB,pred}})^2.
\end{equation}
This corresponds to a squared $L^2$ norm in logarithmic space, emphasizing relative differences and balancing contributions across the dynamic range. While alternative weightings (e.g., asinh) are possible, logarithmic scaling provides a simple and effective choice.

In addition, we optimize the brightness ratio $R_\text{B}= I_\text{pB}/ I_\text{tB}$ between observed and predicted observations
\begin{equation}
L_\text{R} = (R_\text{B, obs} - R_\text{B, pred})^2.
\end{equation}
While this approach can provide tomographic reconstructions of the electron density in the heliosphere, a limited number of viewpoints can result in unconstrained solutions.

\subsubsection{Physics-based Loss Functions}
To further constrain the inversion, we impose a physics-informed loss.
We randomly sample spatiotemporal coordinates $(x,y,z,t)$ within the bounds of the reconstruction volume (i.e., the modeled spherical shell over time) and use them as query points for the same neural network that represents the tomographic model, which maps them to the corresponding plasma quantities.
From the resulting outputs ($\ln{\rho}, \vec{v}$) we compute derivatives with respect to the input coordinates ($\frac{\partial{\ln{\rho}}}{\partial{t}}$, $\frac{\partial{v_x}}{\partial{x}}$, 
$\frac{\partial{v_x}}{\partial{y}}$, ...) and evaluate the continuity equation, which can be written in logarithmic form as
\begin{equation}
\frac{\partial\rho}{\partial t} + \nabla \cdot \left(\rho \vec{v}\right) = 0
\end{equation}
\begin{equation}
\frac{\partial \ln \rho}{\partial t} + \nabla \cdot \vec{v} + \vec{v} \cdot \nabla \ln \rho = 0 \,.
\end{equation}
The squared residuals of the continuity constraint
\begin{equation}
L_C = \left(\frac{\partial \ln \rho}{\partial t} + \nabla \cdot \vec{v} + \vec{v} \cdot \nabla \ln \rho \right)^2 \,,    
\end{equation}
are minimized jointly with the image-based reconstruction, enforcing consistency with the continuity equation throughout the probed volume. This has two key implications. First, the velocity profile is coupled with the dynamics of the reconstructed plasma density. Second, the plasma can only enter through the inner boundary and leave through the outer boundary, while other sinks or sources of plasma density within the spherical shell are suppressed.

In addition, we regularize the solar wind distribution to be approximately radial
\begin{equation}
L_\text{radial} = \left(\frac{\left\|\vec{v} \times \vec{r} \right\|}{\left\|\vec{v}\right\| \left\|\vec{r}\right\|}\right)^2,
\end{equation}
and to be within an appropriate solar wind speed range
\begin{equation}
L_{\text{velocity}} =
\begin{cases}
(\|\vec{v}\| - v_{\min})^2 & \text{if } \|\vec{v}\| < v_{\min} \\
(\|\vec{v}\| - v_{\max})^2 & \text{if } \|\vec{v}\| > v_{\max} \\
0 & \text{otherwise}
\end{cases}
\end{equation}
We set $v_{\min}=200$ and $v_{\max}=800~\mathrm{km\,s^{-1}}$, corresponding to a broad range of typical solar-wind speeds.

With the additional velocity regularization, we aim to mitigate unphysical solutions, such as ghost trajectories that deviate strongly from the radial condition or static plasma clouds that violate the minimum velocity constraint. These limits are chosen such that regular CME dynamics, including deformations, high velocities, and moderate non-radial propagation, are not suppressed. For the reconstructions considered here, the maximum velocity is typically not reached, but the upper bound could be increased for applications closer to the Sun or in the presence of particularly fast CMEs.

\subsubsection{Combined Loss Function}

The combined loss is then given by
\begin{align}
    \label{eq:loss}
    L_\text{total} = & \; \lambda_\text{image} \cdot (L_\mathrm{tB} + L_\mathrm{pB}) \nonumber \\
    & + \lambda_\text{ratio} \cdot L_R \nonumber \\
    & + \lambda_\text{continuity} \cdot L_C \nonumber \\
    & + \lambda_\text{radial} \cdot L_\text{radial} \nonumber \\
    & + \lambda_\text{velocity} \cdot L_\text{velocity} \,,
\end{align}
where the $\lambda$ denote the individual weighting parameters. Note that in the absence of polarized brightness observations this optimization is reduced to only total brightness.

The loss terms are formulated as squared residuals (i.e., $L^2$-type losses), with the image-based terms ($L_{\mathrm{tB}}, L_{\mathrm{pB}}$) evaluated in logarithmic space and the continuity constraint formulated in terms of $\ln \rho$. The consistent use of squared residuals strongly penalize large residuals, while the logarithmic formulation enables coherent optimization across the orders-of-magnitude dynamic range of densities and white-light observations.

\subsubsection{Training Parameters}
Throughout this work we set
$\lambda_\text{image}=1$, $\lambda_\text{ratio}=1$,
$\lambda_\text{continuity}=0.01$, $\lambda_\text{radial}=0.01$,
and $\lambda_\text{velocity}=0.01$.
These values are chosen such that the observational data terms and the continuity equation constitute the primary minimization objectives, while $L_\text{radial}$ and $L_\text{velocity}$ act only as regularization terms.

A systematic ablation study of the physics-informed weighting factors is provided in Appendix~\ref{sec:physics_ablation}, where we demonstrate the existence of a well-defined stability regime. Moderate weights in the range $10^{-2}$–$10^{-1}$ yield optimal performance, whereas excessively large weights lead to degraded reconstructions as the optimization becomes dominated by the physical constraints.
In Sect.~\ref{sec:physics_constraints} we additionally provide a comparison with a pure NeRF approach ($\lambda_\text{continuity}=\lambda_\text{radial}=\lambda_\text{velocity}=0$). 

\subsubsection{Optimization}
The reconstruction is obtained by minimizing the total loss function defined in Eq \ref{eq:loss} using the Adam optimizer. We use an initial learning rate of $10^{-4}$, which we exponentially decay to $10^{-5}$ over $10^{6}$ training iterations.
The number of sampled rays and randomly sampled points can be adjusted according to the available compute resources. For our reconstructions, we use four NVIDIA A100 GPUs, sampling 8192 rays (pixels) for ray tracing and 4096 randomly chosen coordinate points to evaluate the physical constraints. Each model is trained until convergence, which corresponds to approximately 40 epochs and is typically achieved within 12 hours. 

The optimization problem is nonconvex, and convergence is therefore to a local minimum. In contrast to classical tomography methods, reconstruction quality is assessed through synthetic truth tests and robustness analyses rather than through theoretical convergence guarantees.

\subsection{Data}
\label{sec:method_data}

We build on synthetic white-light observations obtained from magnetohydrodynamic simulations of the inner heliosphere with a propagating magnetized CME, performed using the GAMERA-Helio model \citep{Provornikova2024gamera}. In these simulations, the CME structure is represented by the Gibson-Low model \citep{gibson1998}. The solar wind background in GAMERA-Helio is driven by inner boundary conditions obtained from a semi-empirical model of the solar corona, WSA-ADAPT \citep{arge2004stream}. Here, the simulated electron density was used to render synthetic total and polarized brightness observations as described in Sect.~\ref{sec:tomography}. Specifically, we use the public CME challenge dataset\footnote{CME challenge: \url{https://download.hao.ucar.edu/pub/punch/cme_challenge_v2/}} which aims to provide a dataset of synthetic observations and the corresponding ground-truth electron density distribution, as test set for the Polarimeter to Unify the Corona and Heliosphere mission \citep[PUNCH;][]{deforest2022punch, deforest2025PUNCH}. The simulation covers the inner heliospheric domain from 21.5 to 220 $R_\odot$, and -72$^\circ$ to 72$^\circ$ latitude, where the plasma density is specified at the cell centers of the computational grid. 
Therefore, the field of view of the rendered observations is comparable to that from the composite images produced by the three PUNCH/WFI spacecrafts. The observations provide a field of view of 90$^\circ$ and are placed at a distance of 1 AU, however, we truncate the observations to 21.5 - 130 $R_\odot$ in the plane of sky projection. For all our evaluations, we only compare the region between 30 and 120 $R_\odot$. We use CME0 ('reference case') for our study, where the primary propagation direction is in the ecliptic plane (latitude=0$^\circ$). For the observer location, we use the Heliographic latitude and longitude. In this frame, the primary CME propagation direction corresponds to longitude=135$^\circ$. For our reconstructions, we throughout use all available 75 timesteps, which correspond to an observing series of $\sim$1.55 days. In the CME0 simulation, the WSA-ADAPT solution for Carrington Rotation 2095 (27 March - 22 April 2010), corresponding to the magnetogram 26 March 2010 20:00 UT, was used to set the inner boundary conditions.  The defined start time of the simulation is 3 April 2010 09:04 UT. The synthetic observation series ranges from 13 April 2010 19:40 UT to 15 April 2010 08:45 UT, where the first frame corresponds to 250.6 hours of simulation time. Note that the initial 200 hours serve to simulate the solar wind background, into which the CME is subsequently injected. Examples of synthetic $tB$ and $pB$ observations can be seen in Fig. \ref{fig:example}.


Synthetic white-light observations are produced by forward-modeling Thomson-scattered emission from the GAMERA electron density across multiple viewpoints. We use positions at three latitudes ($-40^\circ$, $0^\circ$, $40^\circ$) and longitudes spaced every $20^\circ$. To evaluate our method, we use subsets of these synthetic observations to assess performance under sparse observational coverage.

In Sect.~\ref{sec:noise} we provide an additional comparison of model performance under different noise levels, where we apply additive Gaussian noise to the synthetic observations,
\begin{equation}
    I_{\mathrm{tB}}^{\mathrm{noisy}}(\vec{x})
    = I_{\mathrm{tB}}(\vec{x})
    + \overline{I}\,\mathcal{N}\!\left(0,\sigma^{2}\right),
\end{equation}
\begin{equation}
    I_{\mathrm{pB}}^{\mathrm{noisy}}(\vec{x})
    = I_{\mathrm{pB}}(\vec{x})
    + \overline{I}\,\mathcal{N}\!\left(0,\sigma^{2}\right).
\end{equation}
using noise amplitudes
\(\sigma \in \{0,\,0.01,\,0.05,\,0.1,\,0.2,\,0.3,\,0.4\}\), which are specified as fractions of the mean brightness \(\overline{I}\). After the addition of noise, we mask negative values from further processing.
The corresponding mean signal-to-noise ratio is
\begin{equation}
    \overline{\mathrm{SNR}} = \frac{1}{\sigma}.
\end{equation}
Note that the local noise level varies across the field of view, as the brightness rapidly decreases from the inner to the outer parts of the observations.

\subsection{Evaluation}
\label{sec:evaluation}

For all evaluations, we compute the grid-representation of our reconstruction by sampling the volume in spherical coordinates from 30 to 120 $R_\odot$ in radius. In addition, we unnormalize the model units of our reconstruction to physical units. We compare longitudinal and latitudinal slices of the ground-truth and SuNeRF-CME reconstructed electron density ($\rho$ in $\mathrm{N_e\,cm^{-3}}$), where we use the primary axis of CME propagation (latitude=0$^\circ$, longitude=135$^\circ$). Our quantitative evaluations compare differences per grid-cell, where the polar regions are excluded as given by the GAMERA-Helio simulations. We use the snapshots 10 (14 April 2010 00:41 UT) to 39 (14 April 2010 15:13 UT) for our model evaluation, which corresponds to the time frame from the first clear CME signature to the point where the shock front reaches $\sim100$~$R_\odot$. Note that we use all available time steps for our reconstructions to obtain a better reconstruction of the background solar wind, but limit our evaluation to the duration of the primary CME propagation.

\subsubsection{CME parameters}
\label{sec:cme_extraction}

For determining the primary CME parameters, we first remove the background from both the reconstructed and ground-truth electron density. For this we compute the mean electron density over the first seven snapshots, which correspond to the time period prior to the CME. The obtained background electron density is then subtracted from each individual data volume, resulting in a more isolated CME. All negative values are ignored for further evaluation. Note that this step is performed independently for the reconstructions and the ground-truth data cubes.

From the resulting electron density, we compute the center of mass (CoM) as
\begin{equation}
\label{eq:center_of_mass}
\vec{r}_\text{CoM} = \frac{\sum_{i=1}^N \rho_i \vec{r}_i A_i , dr_i}{\sum_{i=1}^N \rho_i A_i , dr_i}\,,
\end{equation}
where $i$ indexes the grid cells and $N$ is the total number of cells. Here, $\rho_i$ is the electron density at cell $i$, $\vec{r}_i$ the position vector, and $dr_i$ the radial grid spacing at that location. The area element $A_i$ in spherical coordinates is given by $A_i = r_i^2 \sin{\theta_i}$, with $r_i$ and $\theta_i$ denoting the radial distance and latitude, respectively.
From the center of mass, we compute the latitudinal $\theta$ and longitudinal $\phi$ position and determine the difference between the ground-truth (GT) and reconstruction (SuNeRF)($\Delta\theta=|\theta_\text{CoM; GT} - \theta_\text{CoM; SuNeRF}|$, $\Delta\phi=|\phi_\text{CoM; GT} - \phi_\text{CoM; SuNeRF}|$). Analogously to Equation \ref{eq:center_of_mass} we compute the total mass $\sum_{i=1}^N \rho_i A_i , dr_i$, where we use the electron density without background subtraction.
In addition, we determine the position of the shock front, by computing the point with the strongest negative gradient along the primary CME propagation direction (latitude=0$^\circ$, longitude=135$^\circ$).
We estimate the velocity of the center of mass $v_\text{CoM}$ and shock front $v_\text{FRT}$ by applying a linear fit to the radial distance as a function of time.

\subsubsection{Metrics}
\label{sec:metrics}

For the quantitative evaluation, we compare our reconstructed electron density to the corresponding GAMERA simulation snapshots at each time step. For each snapshot, we compute the mean absolute error (MAE) and the cross-correlation coefficient (CC) between the predicted and ground-truth densities. The MAE is defined as
\begin{equation}
\text{MAE} = \frac{1}{N} \sum_{i=1}^N \left| \rho_{\text{SuNeRF}, i} - \rho_{\text{GT}, i} \right|,
\end{equation}
where \( \rho_{\text{SuNeRF}, i} \) and \( \rho_{\text{GT}, i} \) denote the reconstructed and ground-truth electron density at grid point \( i \), respectively, and \( N \) is the total number of grid points.

The cross-correlation coefficient is given by
\begin{equation}
\text{CC} = \frac{\mathrm{Cov}(\rho_{\text{SuNeRF}}, \rho_{\text{GT}})}{\sigma_{\text{SuNeRF}} \, \sigma_{\text{GT}}},
\end{equation}
where \( \mathrm{Cov}(\rho_{\text{SuNeRF}}, \rho_{\text{GT}}) \) is the sample covariance between the reconstructed and ground-truth electron densities, and \( \sigma_{\text{SuNeRF}} \) and \( \sigma_{\text{GT}} \) are their respective standard deviations.

\section{Results}
\label{sec:results}

We use the synthetic white-light observations described in Sect. \ref{sec:method_data} to study the performance of our approach for different observer configurations. In Sect. \ref{sec:heliosphere}, we reconstruct the 3D plasma distribution in the heliosphere based on a variable number of viewpoints. Section \ref{sec:cme_parameters} provides an application to space-weather forecasting, where we compare estimated CME parameters and their viewpoint dependence. To advance this further, we assess the ability of our method to reconstruct the 3D plasma distribution of CMEs (Sect. \ref{sec:results_tomography}). Finally, we evaluate the impact of physics constraints (Sect.~\ref{sec:physics_constraints}), polarization (Sect.~\ref{sec:polarization}), and noise (Sect.~\ref{sec:noise}) on the resulting reconstructions. For all reconstructions, the ground truth is used exclusively for post hoc evaluation of the reconstructions and is not available to, or used by, the reconstruction algorithm itself. We provide reconstructions of additional CME events in Supplementary Sect. \ref{sec:additional_reconstructions}, where all evaluations against the ground truth are performed strictly after training.

\begin{figure}[ht]
    \centering
    \includegraphics[width=\linewidth]{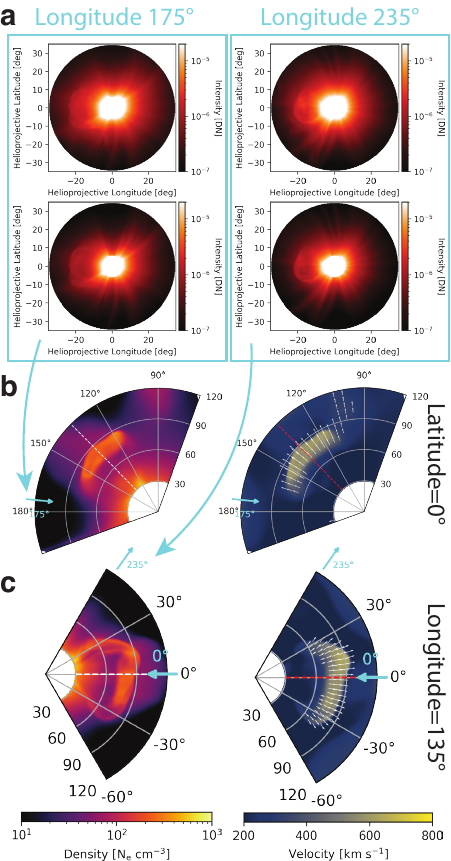}
    \caption{Synthetic observations and corresponding tomographic reconstructions of a CME event, shown 12.5 hours after launch. (a) Synthetic white-light images from two ecliptic viewpoints located at 175$^\circ$ and 235$^\circ$ longitude. (b,c) Slices of the reconstructed electron density and plasma velocity obtained using our tomographic reconstruction method, illustrating the 3D CME structure.} (b) shows ecliptic slices at 0$^\circ$ latitude and (c) slices at 135$^\circ$ longitude. White arrows indicate velocity vectors $>300$~$\mathrm{km\,s^{-1}}$. The white dashed lines mark the main propagation direction of the CME at 0$^\circ$ latitude and 135$^\circ$ longitude in heliocentric coordinates.
    \label{fig:example}
\end{figure}

Figure \ref{fig:example} shows sample input data from two viewpoints, both in terms of total (top) and polarized (bottom) brightness. From the corresponding 3D reconstructions, we extract a longitudinal slice from the ecliptic plane (latitude=0$^\circ$; panel b) and latitudinal slice in the CME propagation direction (longitude=135$^\circ$; panel c). In addition, our approach intrinsically models the plasma velocity, by minimizing the continuity equation. The derived velocity maps (panel b and c) show the increased velocity of the CME plasma ($\sim$600~$\mathrm{km\,s^{-1}}$), and the background solar wind speed of $\sim$250~$\mathrm{km\,s^{-1}}$. While the velocity field can provide additional insights into the CME dynamics, we focus our further evaluation on the reconstructed plasma density only.

\subsection{Heliospheric reconstruction}
\label{sec:heliosphere}

\begin{figure*}[ht]
    \centering
    \includegraphics[width=\linewidth]{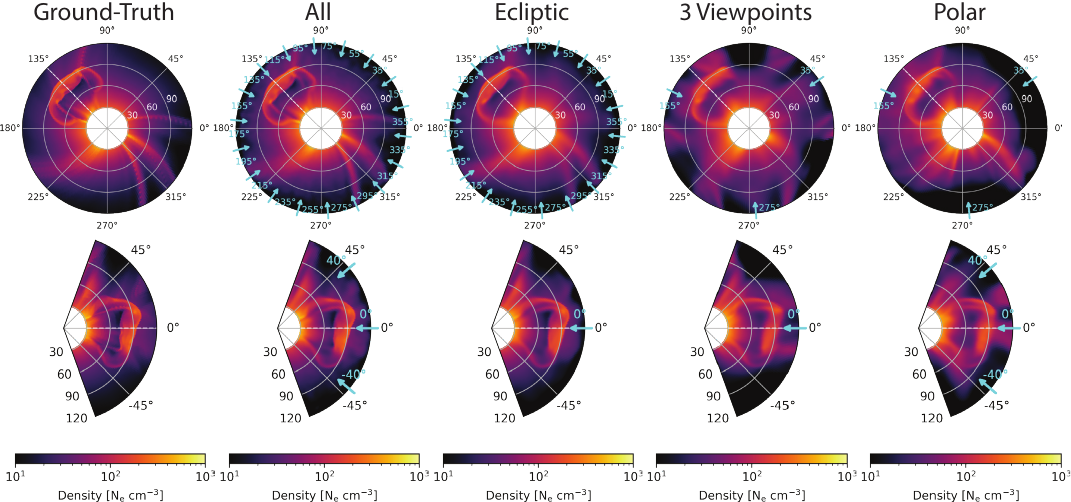}
    \caption{Tomographic reconstruction of the full heliosphere from varying observer configurations. We compare ecliptic (top) and latitudinal slices at 135$^\circ$ longitude (bottom) for the ground truth and reconstructions using all available viewpoints, only ecliptic viewpoints, three ecliptic-plane observers, and a high-latitude configuration (polar). Observer positions are indicated by blue arrows. The results show that additional viewpoints improve reconstruction quality, particularly for the background solar wind. Reconstructions are more accurate when the target structures are viewed from multiple angles (latitudinal slices), while reconstructions within the observer plane are less reliable.}
    \label{fig:heliospheric_mapping}
\end{figure*}

We apply our approach to a varying number of input observations and compare the 3D reconstructions of plasma density to the ground-truth data. Here, we compare reconstructions with the full set of synthetic observations (total of 54 viewpoints; ``All''), observations from the ecliptic plane only (total of 19 viewpoints; ``Ecliptic''), to three simultaneously and equally spaced observers (``3 Views''), and to a configuration of three observers at different latitudes (-40, 0 , 40$^\circ$ latitude; ``Polar''). Figure \ref{fig:heliospheric_mapping} shows a direct comparison of the latitudinal and longitudinal slices through the reconstructed 3D volume. The configuration with all observers demonstrates that the method can directly integrate all available observations into a unified 3D picture of the electron density, where even small-scale streamers in the ecliptic plane are reconstructed. In contrast, reconstructions using only ecliptic viewpoints result in a noticeably coarser representation in the ecliptic plane. The comparison to the longitudinal slice shows that similar reconstructions are obtained out of the ecliptic. This demonstrates that reconstructions in the observer plane are typically less reliable than those from out-of-plane (i.e., at higher latitudes). The configuration with three viewpoints demonstrates a realistic setting (e.g., SOHO/LASCO + STEREO-A/COR + STEREO-B/COR) and shows again that reconstructions in the observer plane are typically less reliable. In addition, the reduced number of observers makes the reconstruction of solar wind streamers more challenging. However, the CME propagation direction, extent and position of the shock front is well captured, in particular in the longitudinal slice. As a theoretical configuration, we also examine the configuration with observers at multiple latitudes, which could be provided by a future solar polar mission. While this configuration also only uses three viewpoints, the slices demonstrate an improvement of the reconstructed solar wind structures in the ecliptic plane.

Table \ref{table:heliospheric_mapping} provides a quantitative evaluation of the reconstruction, where we compute the correlation coefficients and the absolute error of $\rho$ in $\mathrm{N_e\,cm^{-3}}$ and in relative units. We compute the metrics per snapshot within -60$^\circ$ to 60$^\circ$ in latitude and then compute the mean value and standard deviation over the full set of 30 snapshots. Throughout, a larger number of viewpoints leads to improved reconstructions, where ``All'' and ``Ecliptic'' achieve correlation coefficients of 0.99 and 0.98, respectively. The realistic configuration with three observers shows a stronger decrease in correlation to 0.93, and absolute errors in the range 26\%. This is largely attributed to the background solar wind structure, which shows increased deviations from the ground-truth. The comparison to the ``Polar'' configuration shows that on average a similar performance is achieved (e.g., correlation of 0.92). This indicates that while observations from higher latitudes can lead to better reconstructions in the ecliptic plane, the full heliospheric reconstruction is primarily determined by the number of total viewpoints. We note that the cross-correlation coefficient can remain relatively high even when background structures are not perfectly captured, owing to the dominant radial density gradient. The high correlation coefficient is largely influenced by the dominant radial density gradient, while improvements in the reconstructions are more clearly reflected in the relative variation.

\begin{table}[h!]
\centering
\caption{Comparison of methods based on density correlation coefficient, MAE, and relative MAE}
\begin{tabular}{|l|c|c|c|}
\hline
\textbf{Method} & \textbf{CC} & \textbf{MAE [$\mathrm{N_e\,cm^{-3}}$]} & \textbf{MAE [\%]} \\
\hline
All & $0.99 \pm 0.00$ & $3.32 \pm 0.06$ & $4.97 \pm 0.05$ \\
Ecliptic & $0.98 \pm 0.00$ & $5.53 \pm 0.15$ & $8.29 \pm 0.15$ \\
3 Views & $0.93 \pm 0.00$ & $17.19 \pm 0.93$ & $25.75 \pm 1.17$ \\
Polar & $0.92 \pm 0.00$ & $17.37 \pm 0.74$ & $26.03 \pm 0.88$ \\
\hline
\end{tabular}
\label{table:heliospheric_mapping}
\end{table}

\subsection{Estimation of CME parameters}
\label{sec:cme_parameters}

\begin{figure*}[t]
    \centering
    \includegraphics[width=\linewidth]{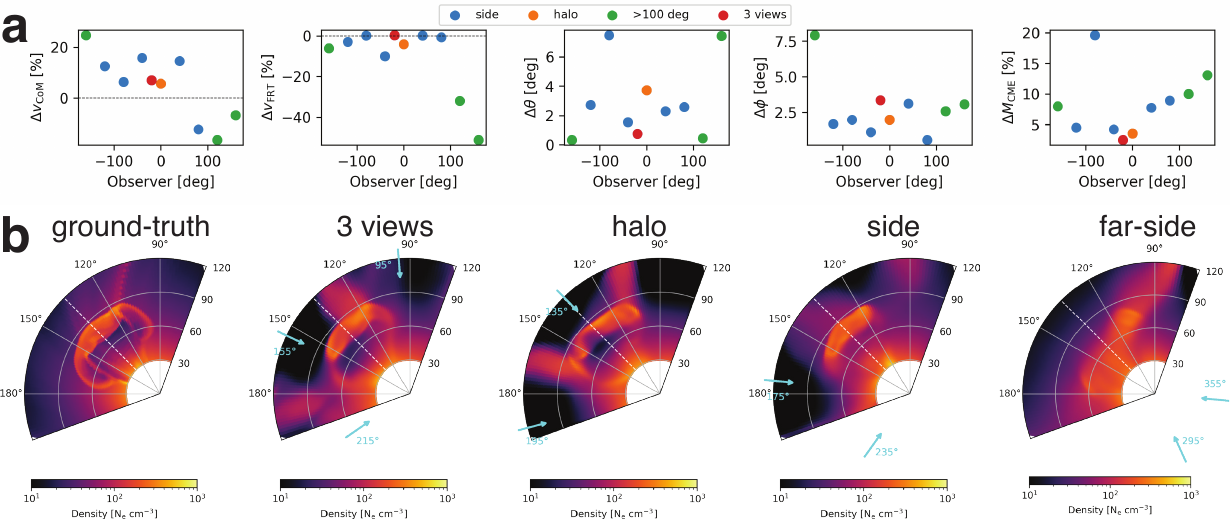}
    \caption{Evaluation of derived CME parameters across different observer configurations.
    (a) Errors in center-of-mass velocity (\(\Delta v_\text{CoM}\)), shock front velocity (\(\Delta v_\text{FRT}\)), latitudinal position (\(\Delta\theta\)), longitudinal position (\(\Delta\phi\)), and total CME mass (\(\Delta M_\text{CME}\)).
    (b) Ecliptic slices of the reconstructed electron density for the ground truth, a three-viewpoint configuration, a halo-CME setup, two side-on viewpoints, and a far-side configuration. Observer positions are indicated by blue arrows.
    The most accurate reconstructions are obtained when observers are located within 100$^\circ$ of the CME. However, the propagation direction can still be reasonably recovered even from far-side observations.}
    \label{fig:cme_parameters}
\end{figure*}

From a space-weather perspective it is of primary importance to estimate CME parameters, in terms of propagation direction, velocity, and total mass. In this section, we compute the CME parameters from the 3D electron density both from the ground-truth and the tomographic reconstructions. We use configurations with two observers with a 60$^\circ$ separation angle and study their viewpoint dependency in the range from 0$^\circ$ (halo) to 320$^\circ$ in steps of 40$^\circ$ longitude. This is a representative configuration that could be achieved with the anticipated Vigil mission \citep{kraft2017vigil}, located at Lagrange point 5, in combination with observations from PUNCH or the Large Angle Spectroscopic Coronagraph \citep[LASCO; ][]{Brueckner1995lasco}. In addition, we compare a constellation of three observers that represent observations from Lagrange points 1, 4, and 5, which could be achieved through the addition of the STEREO mission \citep{kaiser2008stereo}. Note that we consider idealized observations with the same field of view and calibration, while in practice observational differences can impose additional challenges.

As described in Sect. \ref{sec:cme_extraction}, we apply a background subtraction and estimate the center of mass. We compute the average velocity, latitudinal direction, and longitudinal direction over the CME evolution. In addition, we estimate the speed of the shock front in the primary propagation direction. The mass difference is computed from the entire volume spanning $-60^\circ$ to $60^\circ$ in latitude, $70^\circ$ to $200^\circ$ in longitude, and 30 to 120~$R_\odot$ in radial distance from the solar center. Note that we explicitly refrain from  using the actual CME parameters that were used as input for the GAMERA simulation to provide a more independent evaluation that also includes the background solar wind structure.

Figure~\ref{fig:cme_parameters} provides a comparison between the parameters derived from our tomographic reconstructions and those obtained from the ground-truth plasma distribution (panel a). The angular distance shown on the horizontal axis is defined with respect to the CME propagation direction, where $0^\circ$ corresponds to a full halo CME as seen from the observer's viewpoint.
Panel b displays longitudinal slices of the electron density in the ecliptic plane, comparing the ground truth (left), the reconstruction using three viewpoints (center left), and reconstructions using three different two-viewpoint configurations (right panels). Observer locations are indicated by blue arrows. The two-viewpoint cases include: a full-halo CME configuration, a side view with an $80^\circ$ angular offset, and a far-side reconstruction with a $160^\circ$ offset.

Among all configurations, the three-viewpoint reconstruction yields the best parameter estimates. For the cases with two viewpoints where at least one observer is within $100^\circ$ of the CME propagation direction, we also achieve reliable CME parameter estimates. Across these configurations, we obtain the following mean errors with respect to the ground truth:
\begin{itemize}
    \item Latitude: 3.39 $\pm$ 1.94 $^\circ$
    \item Longitude: 1.76 $\pm$ 0.79 $^\circ$
    \item Shock front velocity: 3.01 $\pm$ 3.45 \%
    \item Center-of-mass (CoM) velocity: 11.24 $\pm$ 3.87 \%
    \item Total mass: 8.10 $\pm$ 5.50 \%
\end{itemize}

These results demonstrate that our method can reliably recover CME kinematic and geometric parameters across a broad range of observer configurations, provided that the viewing geometry ensures favorable line-of-sight intersection with the CME structure. A noticeable reduction in performance occurs only when one of the observers is located on the solar far side, with an angular separation $>160^\circ$ in longitude from the CME propagation direction. In these cases, the reconstructions show increased uncertainties, with velocity deviations $>20\%$ and latitude errors $>5^\circ$.
Even under this far-side configuration, where the two observers are positioned at $160^\circ$ and $-140^\circ$ relative to the propagation direction, the model recovers the CME trajectory with longitude and latitude errors below $<8^\circ$ and $<9^\circ$, respectively.
We also note a slight bias in parameter estimation toward the center of mass, introduced by the centered evaluation volume. This choice ensures consistent comparisons across configurations and avoids contamination from regions outside the valid reconstruction domain.

The velocity estimates of the center of mass are in the range of 10\%, while velocity estimates of the shock-front are in the range of 5\% (excluding far-side configurations). This shows that the shock front can be well separated from the background solar wind and appears as distinct density variation in our reconstructions. This is in agreement with related reconstruction methods that are based on polarized light, where the best reconstructions are expected for confined regions of increased electron density \citep{deKonig2011polarimetric_localization, Gibson2025PolarizationDiagnostics}.

Similar to the full heliospheric reconstruction (Sect. \ref{sec:heliosphere}), a primary limitation is the approximation of the background structures. Here, a longer time sequence or more sophisticated background estimation could help to disentangle the CME from the more static solar wind structures.

\subsection{CME tomography}
\label{sec:results_tomography}

\begin{figure*}[ht]
    \centering
    \includegraphics[width=0.95\linewidth]{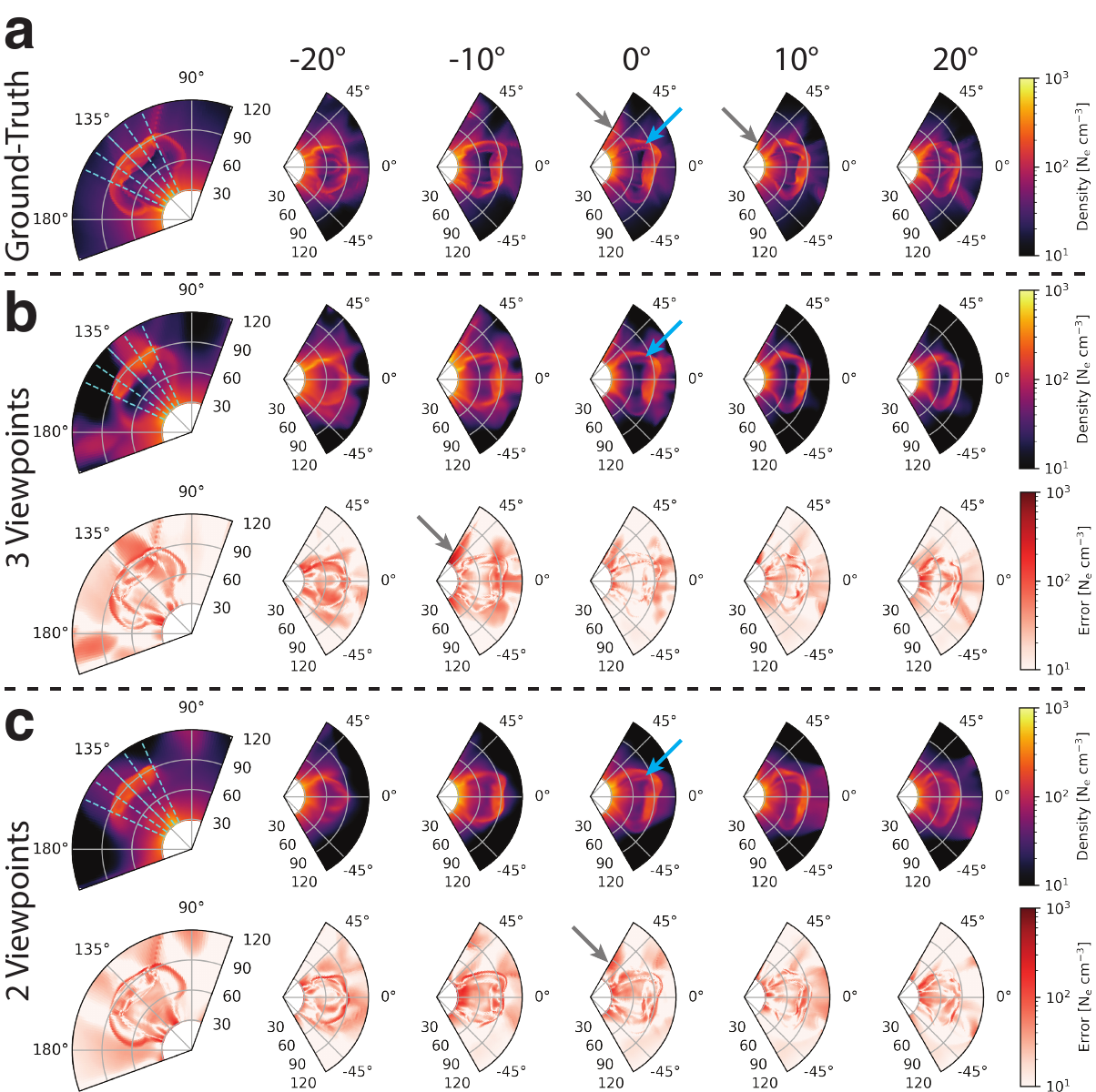}
    \caption{Tomographic reconstruction of a CME from three and two viewpoints.
    (a) Ground-truth electron density slices in the ecliptic plane and at 135$^\circ$ longitude; the blue dashed lines indicate the locations of the latitudinal slices.
    (b) Reconstructed slices from a three-viewpoint configuration and corresponding difference maps relative to the ground truth.
    (c) Same as (b), but for a two-viewpoint reconstruction.
    The comparison highlights that the CME morphology can be reconstructed with a high degree of fidelity even from a sparse set of observations.}
    \label{fig:tomography}
\end{figure*}

While the estimation of CME parameters can be a useful input for space-weather forecasting, a more complete understanding of the CME morphology and its temporal evolution is required to understand the physical processes driving plasma propagation in the heliosphere \citep{temmer2023propagation}. In this section, we explore the ability of our method to fill this gap by estimating the 3D structure based on a limited number of observations.

Figure \ref{fig:tomography} provides an analysis of a full tomographic reconstruction from three and two viewpoints only, and compares slices of electron density to the ground-truth data. The first row shows latitudinal slices through the ecliptic plane, where the blue dashed lines with a spacing of 10$^\circ$ indicate the longitude positions from which we extract the slices. Panel a shows the ground-truth distribution of electron density for the respective slices. In panel b, we show the reconstruction results from three viewpoints, where we use the same viewpoint configuration as in Sect. \ref{sec:cme_parameters} (95$^\circ$, 155$^\circ$, 215$^\circ$ in heliographic longitude). Panel c shows the reconstructions with two viewpoints at 175$^\circ$ and 235$^\circ$ in heliographic longitude. In addition to the extracted slices we provide difference maps between our reconstructions and the ground-truth data.
The comparison demonstrates that with only two viewpoints, the model yields a 3D tomographic reconstruction that captures the key CME morphological features, including the bright core, dark cavity, and bright shock front. For both reconstructions, the CME orientation and spatial extent are in good agreement with the actual plasma distribution. Moreover, the reconstructions capture the deformed CME front, where the northern front is slightly more extended in the radial direction. The slices also show that the model reproduces the internal CME structure (blue arrows). The difference maps show that most discrepancies occur near the shock front, where both reconstructions exhibit a slight positional shift that makes the CME appear more compact.
While the CME itself is reconstructed with high fidelity, background structures are more challenging, which can be attributed to the limited number of viewpoints and relatively short time frame. The streamer in Figure \ref{fig:tomography} (gray arrows) appears shifted by  $>10^\circ$ longitude for the reconstruction with three viewpoints and is not recovered in the two-viewpoint case. In general, the reconstructions with three viewpoints show more spatial details and are in better agreement with the ground truth, indicating that our method benefits directly from additional viewpoints. Additional 3D reconstructions of different CMEs from the CME Challenge dataset are provided in Supplementary Sect.~\ref{sec:additional_reconstructions}, including tomographic results obtained from two to four viewpoints.

\subsection{Importance of physics constraints}
\label{sec:physics_constraints}

\begin{figure*}[ht!]
    \centering
    \includegraphics[width=\linewidth]{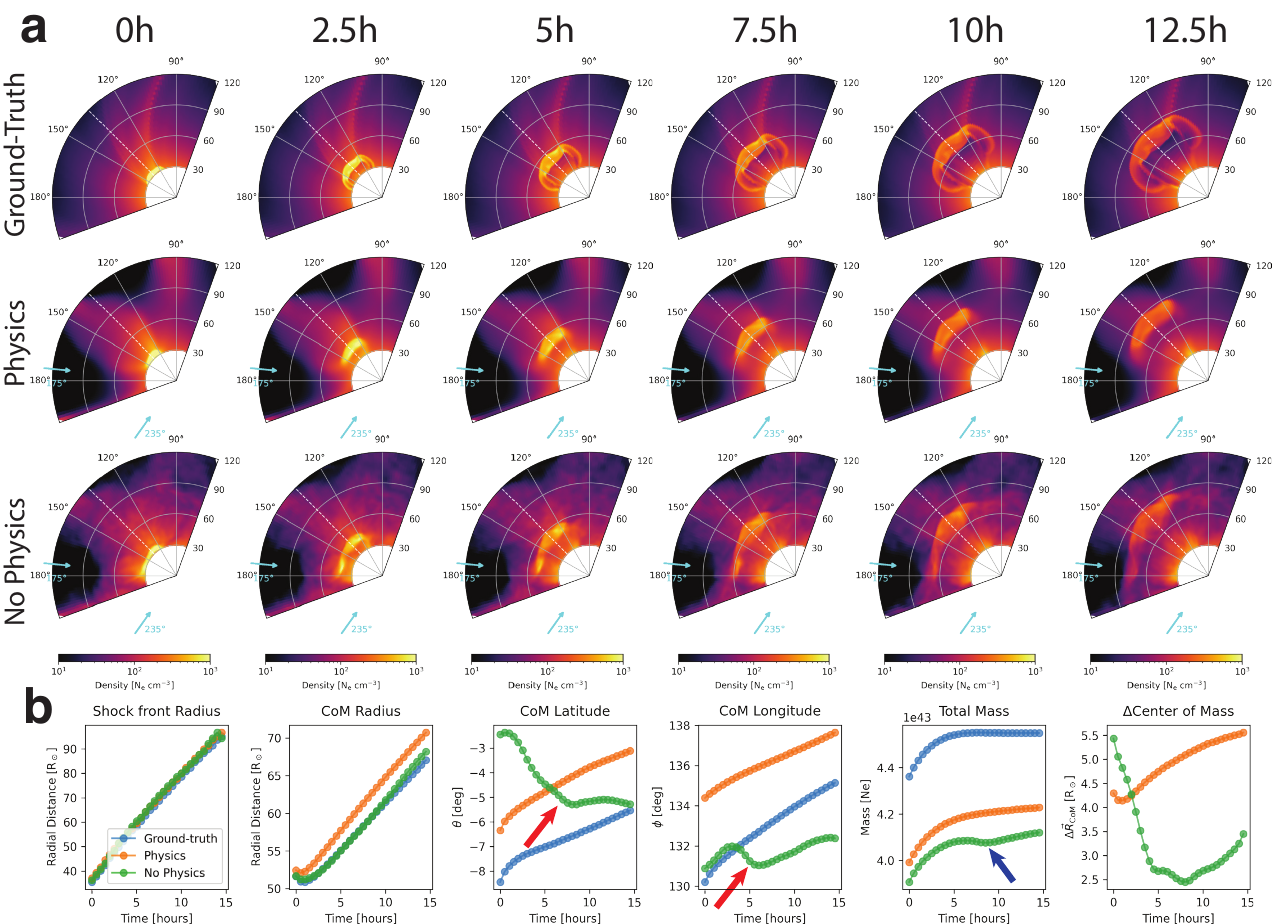}
    \caption{Comparison of tomographic CME reconstructions with and without physics-based constraints.
    (a) Ecliptic slices of electron density over a 12.5-hour interval for the ground truth, the reconstruction with physics-informed losses, and the reconstruction without additional constraints. Both reconstructions are based on observations from two viewpoints, indicated by blue arrows.
    (b) Temporal evolution of key CME parameters, including the radial position of the shock front and center of mass, the latitude and longitude of the center of mass, the total CME mass, and the relative distance between the reconstructed and ground-truth center of mass.
    Including physics-based constraints results in a more physically coherent evolution, reducing abrupt changes in position and total mass.}
    \label{fig:series}
\end{figure*}

In this section, we assess the impact of incorporating additional physical constraints into the tomographic reconstruction. To illustrate their role, we compare two reconstructions: one including the physics-informed loss terms and one omitting them entirely. Our focus is on the temporal evolution of the reconstructed CME, where physical consistency across time is particularly critical. A systematic ablation of the corresponding weighting factors is provided in Appendix~\ref{sec:physics_ablation}

In Fig.~\ref{fig:series}a we compare a series of latitudinal slices through the ecliptic plane over 12.5 hours between the ground truth, the tomographic reconstruction with physical constraints, and the reconstruction without physical constraints. Both reconstructions use two viewpoints at 175$^\circ$ and 235$^\circ$ longitude, as indicated by the blue arrows. The temporal evolution shows that both reconstructions capture the propagation of the CME plasma, however the additional physical constraints provide a more coherent evolution, while the unconstrained reconstructions show additional projection artifacts at around 180$^\circ$ longitude. This can also be seen from panel b where we show the radial distance of the shock front, the spatial location of the center of mass (radius, latitude, longitude), and the total mass as a function of time. Here, the physically constrained reconstructions better reproduce the temporal evolution of the position of the center of mass, while the unconstrained reconstructions show a more random drift of the CME propagation direction (red arrows). For the total mass we notice an offset for both reconstructions, but the overall evolution is well reproduced by the additional physics constraints, which is attributed to the continuity equation, which mitigates sinks or sources of electron density within the modeled volume. In contrast, the unconstrained reconstructions show an unphysical drop in total mass (blue arrow).

We further note that including the continuity equation in the optimization has further implications on the background solar wind structure, where the constrained reconstructions result in a smoother background. While the ground-truth data show a slowly evolving streamer ($\sim100^\circ - 150^\circ$ longitude), both approaches fail to reconstruct the background structures from the limited set of observations.

\subsection{Importance of polarized observations }
\label{sec:polarization}

\begin{figure*}[ht]
    \centering
    \includegraphics[width=\linewidth]{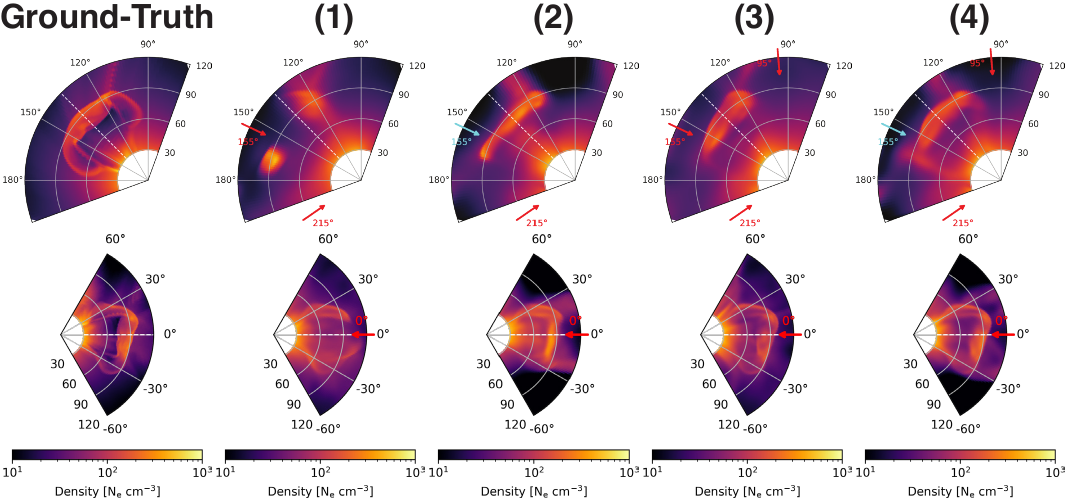}
    \caption{Tomographic reconstructions for different polarization configurations. Ecliptic slices (top) and latitudinal slices at $135^\circ$ longitude (bottom) are shown for the ground truth and the corresponding reconstructions using the configurations described in the text (1-4). Observer positions providing both total and polarized brightness are indicated by blue arrows, while red arrows mark observers with total brightness only. The results demonstrate that including polarized brightness observations substantially improves the reconstruction quality.}
    \label{fig:polarization}
\end{figure*}

The previous evaluation focused on configurations with both polarized and total brightness observations, whereas in practice frequently only total brightness measurements are available. Here, we assess the performance of our reconstruction approach for different combinations of total ($tB$) and polarized ($pB$) brightness data. When only $tB$ observations are available, the loss terms for polarized brightness and the polarization ratio ($L_\text{P}$ and $L_\text{ratio}$) are omitted from the total loss (Eq.~\ref{eq:loss}).

Figure~\ref{fig:polarization} summarizes the tomographic reconstructions obtained from the following configurations:
\begin{enumerate}
    \item two viewpoints with $tB$ observations only;
    \item two viewpoints where one observer provides both $tB$ and $pB$, and the other provides only $tB$;
    \item three viewpoints with $tB$ observations only;
    \item three viewpoints where one observer provides both $tB$ and $pB$.
\end{enumerate}

Slices through the ecliptic plane and along the CME propagation direction show that two viewpoints with $tB$ alone are insufficient to reconstruct the CME. Adding a single $pB$ observation significantly improves the reconstruction, where the primary propagation direction and the overall extent are recovered, although the results still lack spatial detail and the position and extent of the CME front remain limited. The configuration with three $tB$ viewpoints improves the recovered CME extent, but the deformation of the shock front in the ecliptic slice is only partially reproduced. A combined configuration with two $tB$ observers and one observer providing both $tB$ and $pB$ yields reconstructions comparable to those obtained from two viewpoints with both $tB$ and $pB$ (Fig.~\ref{fig:cme_parameters}b). This configuration is also relevant from an observational standpoint, as similar combinations of instruments are available through PUNCH, STEREO, and Parker Solar Probe. This behavior is consistent with the polarization diagnostics of \citet{Gibson2025PolarizationDiagnostics}, which show that $pB$ measurements provide strong constraints on line-of-sight localization through the Thomson scattering geometry. In our tomographic framework, this additional depth information reduces reconstruction ambiguities, such that a single $pB$ viewpoint can effectively substitute for an additional $tB$ viewpoint when combined with multi-view observations. We also note that initial experiments using only a single viewpoint, even when providing both $tB$ and $pB$, did not yield successful reconstructions.

\subsection{Noise robustness }
\label{sec:noise}

While our evaluation is primarily based on idealized synthetic data, we also provide a basic analysis of how noise affects the reconstructions. To this end, we apply gaussian distributed additive noise to each frame, with the noise level~$\sigma$ scaled by the mean brightness of the corresponding frame (see Sect.~\ref{sec:method_data}). 

We emphasize that real coronagraph observations are subject to a more complex combination of noise sources, including Poisson-distributed photon statistics, detector readout noise, amplification effects, and stray-light contamination. Our adopted Gaussian model does not aim to replicate the full instrumental noise characteristics. Instead, it provides a tunable and controlled framework to evaluate reconstruction stability under progressively degraded signal conditions.
Because observational noise levels vary across instruments and observed regions, we perform a broad parameter sweep ranging from noise-free images to strongly degraded cases corresponding to signal-to-noise ratios below 5. This analysis is intended to quantify the intrinsic robustness of the inversion framework rather than to reproduce instrument-specific noise properties.

\begin{figure}[ht]
    \centering
    \includegraphics[width=\linewidth]{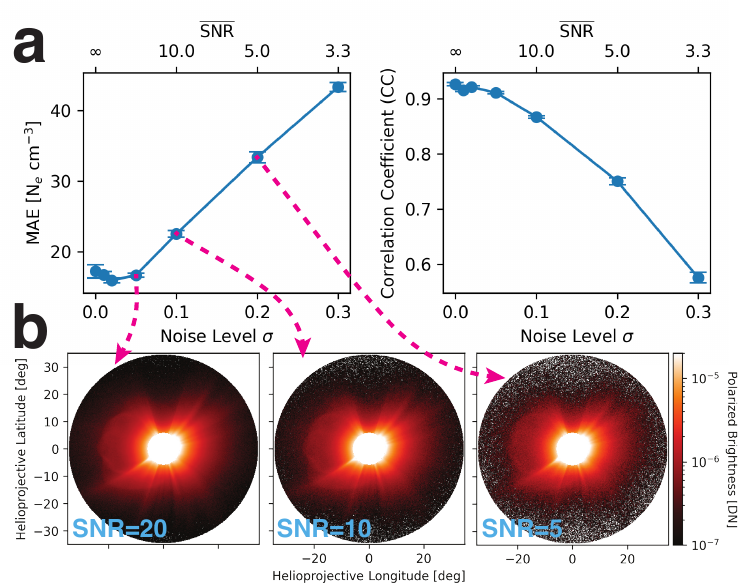}
    \caption{Impact of noise on reconstruction performance. (a) Mean absolute error (MAE) and cross-correlation (CC) of the reconstructions as a function of the applied noise level~$\sigma$, where the noise amplitude is scaled by the mean brightness of each frame. The model remains robust up to noise levels of approximately $5\%$ of the mean brightness ($\overline{\mathrm{SNR}} = 20$), after which the performance gradually degrades. Values represent the mean over the full time series, with error bars indicating the standard deviation. (b) Example of a noisy input frame illustrating that noise becomes dominant in the faint outer regions of the corona, leading to the reduced reconstruction quality observed at higher noise levels.}
    \label{fig:noise}
\end{figure}

Panel~(a) of Fig.~\ref{fig:noise} shows the model performance, quantified by MAE and CC, as a function of the applied noise. The reconstructions remain stable up to noise levels of approximately 5\% of the mean brightness ($\overline{\mathrm{SNR}} = 20$) and then gradually degrade as the noise increases. This degradation is primarily caused by the amplified noise in the faint outer regions of the observations, where the signal becomes comparable to or weaker than the noise (Fig.~\ref{fig:noise}b).

The robustness to noise levels corresponding to $\overline{\mathrm{SNR}} \sim 20$ is encouraging for applications to real observations. Extending the method to observational data will, however, require addressing additional complexities such as varying fields of view, calibration accuracy, stray light, background stars, and other instrumental noise sources. While the framework introduced here provides the core components to tackle these challenges, we defer the modeling of these observational effects to future applications, where they can be incorporated directly into synthetic test data tailored to the specific instrument characteristics.

\section{Discussion}
\label{sec:discussion}

In this study, we presented a method for tomographic reconstructions of electron density in the heliosphere based on observation series of total and polarized brightness. We evaluated the performance of our approach in dependence of different viewpoints using synthetic observations from a GAMERA simulation and comparing the reconstruction result to the corresponding ground-truth electron density. While 3D reconstructions of CMEs from a limited number of viewpoints (e.g., two or three observers) are considered challenging, we showed that our physics-informed neural radiance field can provide tomographic reconstructions of CMEs even from two viewpoints.

All reconstructions in this work are based exclusively on the available observational data and the applied physical constraints, without relying on pre-training or prior knowledge from other events. Each model is trained from scratch for a given CME event, making the reconstructions entirely data-driven and event-specific. 



Our reconstructions demonstrate that the method efficiently integrates available observations, with performance improving as the number of viewpoints increases: from a correlation coefficient (CC) of 0.98 and a mean absolute error (MAE) of 8.3\% for ecliptic-plane configurations to a CC of 0.92 and MAE of 25.8\% in the more realistic three-viewpoint case (Table \ref{table:heliospheric_mapping}). The primary errors stem from background solar wind structure, while CME features remain well recovered, though reconstructions in the observer plane (ecliptic) are less reliable than those from out-of-plane views. For space-weather applications, the approach shows strong potential in estimating CME parameters, achieving mean errors of $3.39\pm1.94^\circ$ in latitude, $1.76\pm0.79^\circ$ in longitude, $3.01\pm3.45\%$ in shock-front velocity, and $8.10\pm5.50\%$ in total mass with two observers separated by $60^\circ$ longitude (Fig. \ref{fig:cme_parameters}). As compared to the performance metrics reported in \citet{2023AdSpR..72.5243V}, this corresponds to an improvement of roughly a factor of two to four over GCS fitting results. Far-side configurations recover approximate propagation directions ($|\Delta\phi|<8^\circ$), though with larger uncertainties. Our method also provides 3D tomographic reconstructions of CMEs, reproducing the three-part structure, deformed shock front, and internal density variations from as few as two viewpoints, while additional observers enhance spatial detail and shock front localization. Using both polarized and total brightness observations yields the most accurate reconstructions, whereas two viewpoints with total brightness alone are insufficient for reliable results. The comparison across noise levels offers an initial assessment of applicability to observational data and indicates robustness up to noise levels of 5\% of the mean brightness ($\overline{\text{SNR}} = 20$), although future work should incorporate additional observational realism (e.g., varying fields of view, background stars, and instrumental effects).

The presented results illustrate the applicability of NeRFs for tomographic reconstructions of astrophysical observations. This method can be flexibly adjusted to the underlying physics of image formation, and smoothly integrate data from multiple viewpoints into a unified 3D reconstruction. In this study, we demonstrated that NeRFs can be directly coupled with PINNs to solve PDEs as part of the model optimization to mitigate unphysical reconstructions. In the present case of CME tomography, the additional physical constraints can help to enforce realistic propagation dynamics, suppressing non-radial or ghost trajectories \citep{deforest2013ghost}. The intrinsic spatio-temporal smoothness of neural representations has further implications on the reconstructions obtained where coherent solutions are intrinsically preferred by the model \citep[c.f.,][]{jarolim2025pinnme, base2025inversion, ramos2023tomography}.

One remaining limitation is the treatment of the ambient solar wind background, which affects CME reconstructions. While additional simultaneous viewpoints are not available, the reconstructions could be improved by incorporating longer temporal sequences. Such an approach is particularly suited for more quasi-stationary solar wind structures and the inner heliosphere, where solar rotation can provide additional constraints to approximate the background and better couple the CME with its surrounding plasma \citep{ramos2023tomography}.

A direct next step is the application to observation data, where this approach has potential to provide a 3D tomographic reconstruction of a CME and its dynamic evolution.
The results presented in this study are based on synthetic observations and a basic noise assessment. Applying the method to real data will require accounting for additional effects such as variable fields of view, scattered light, solar energetic particles, and background stars. The SuNeRF-CME framework is well suited to incorporate these challenges, as noisy inputs can be combined naturally with physical constraints during reconstruction \citep{karniadakis2021physicsml}. The approach also offers the possibility of extending the model to include the full set of MHD equations and an inner boundary condition, enabling heliospheric simulations that are directly informed by coronagraphic observations.

\section{Data availability}
\label{sec:dataavail}

All our 3D reconstruction results and codes are publicly available.
\begin{itemize}
    \item Data (Zenodo): \citet{zenodo_data} 
    \item Data (Globus): \url{https://app.globus.org/file-manager?origin_id=503a88c1-d429-45f8-b2fb-e9ceee5f24ba&origin_path=%2F}
    \item Code: \url{https://github.com/RobertJaro/SuNeRF}
    \item Code (Zenodo): \citet{zenodo_code}
\end{itemize}

\section{Acknowledgments}

RJ was supported by the NASA Jack-Eddy Fellowship and the NASA MDRAIT project (NASA Award Number: 80NSSC23K1025).
This work is a research product of ESL (ESL.ai), an initiative of the Frontier Development Lab, delivered by Trillium Technologies in partnership with ESA, University of Oxford, Google Cloud, and NVIDIA. We gratefully acknowledge Google Cloud, NVIDIA Corporation and ScanAI for providing extensive computational resources.
This material is based upon work supported by the NSF National Center for Atmospheric Research (NCAR), which is a major facility sponsored by the U.S. National Science Foundation under Cooperative Agreement No. 1852977. We would like to acknowledge high-performance computing support from the Derecho system (doi:10.5065/qx9a-pg09) provided by NCAR, sponsored by the National Science Foundation. The CME Challenge dataset was sponsored by NASA 80GSFC18C0014 and DOD-USAF-AOSR FA95502110457 awards.
We would like to thank the external reviewers—Angelos Vourlidas, Manuela Temmer, Erika Palmerio, Benoit Tremblay, and Cooper Downs—for their valuable feedback throughout the ESL sprint.
The authors acknowledge the use of ChatGPT (OpenAI) for language editing and stylistic improvements. The authors are solely responsible for the scientific content, analysis, and conclusions presented in this work.

This research has made use of SunPy \citep{sunpysoftware2020, sunpycommunity2020}, AstroPy \citep{astropy:2013, astropy:2018, astropy:2022}, and PyTorch \citep{pytorch2019_9015}.

%

\vspace{5mm}
\facilities{}


\software{AstroPy \citep{astropy:2013, astropy:2018, astropy:2022},
          SunPy \citep{sunpycommunity2020, sunpysoftware2020},
          PyTorch \citep{pytorch2019_9015}.
          }



\newpage
\appendix

\begin{figure}[!h]
    \centering
    \includegraphics[width=\linewidth]{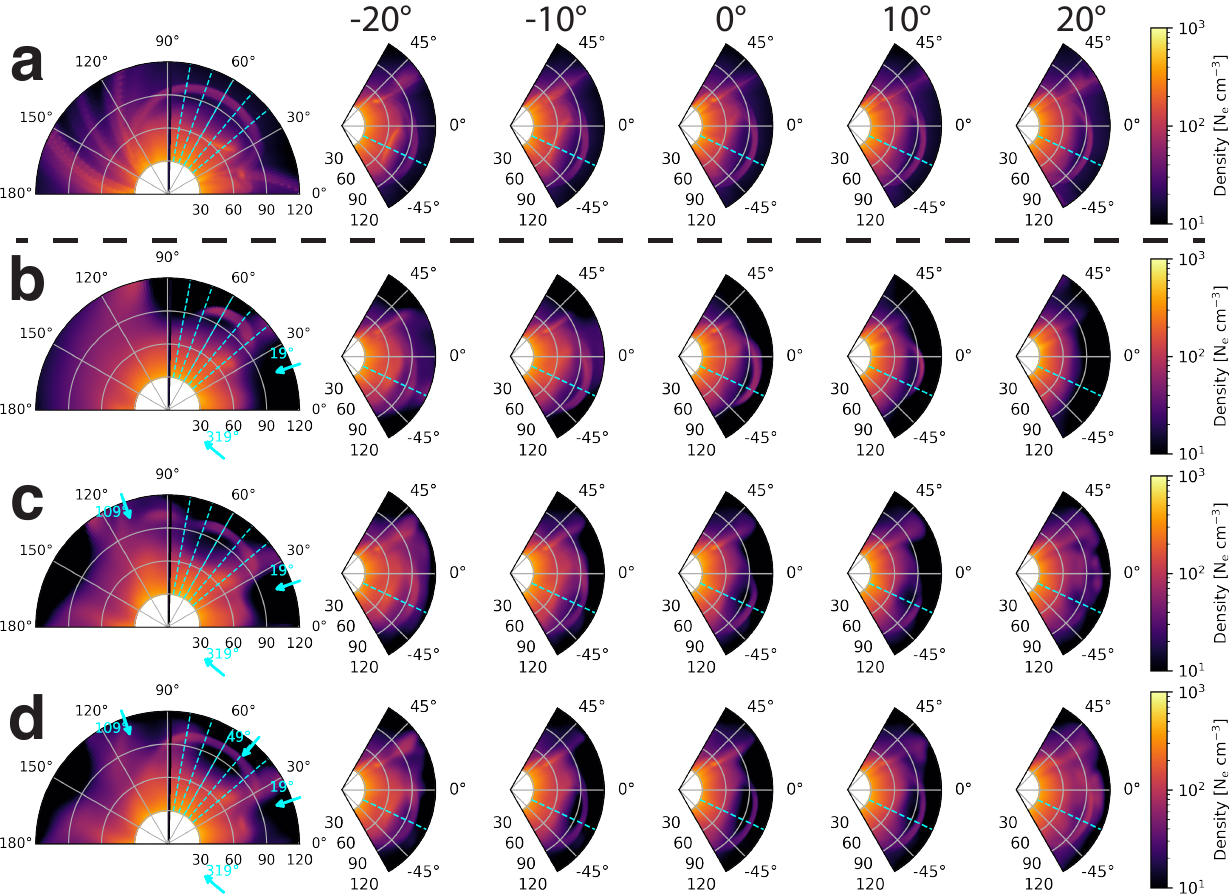}
    \caption{ Tomographic reconstruction of ``CME1'' from the CME Challenge.
(a) Ground-truth electron-density slices at $-25^\circ$ latitude and $60^\circ$ longitude; the blue dashed lines mark the slice locations ($-20^\circ$, $-10^\circ$, $0^\circ$, $10^\circ$, $20^\circ$ relative to central longitude).
(b–d) Reconstructed slices obtained from two-, three-, and four-viewpoint configurations. Observer locations are indicated by blue arrows. Note that the latitude slices are not plane-parallel but correspond to conical surfaces at constant latitude.}
    \label{fig:cme1}
\end{figure}

\section{Additional reconstructions}
\label{sec:additional_reconstructions}

\begin{figure}
    \centering
    \includegraphics[width=\linewidth]{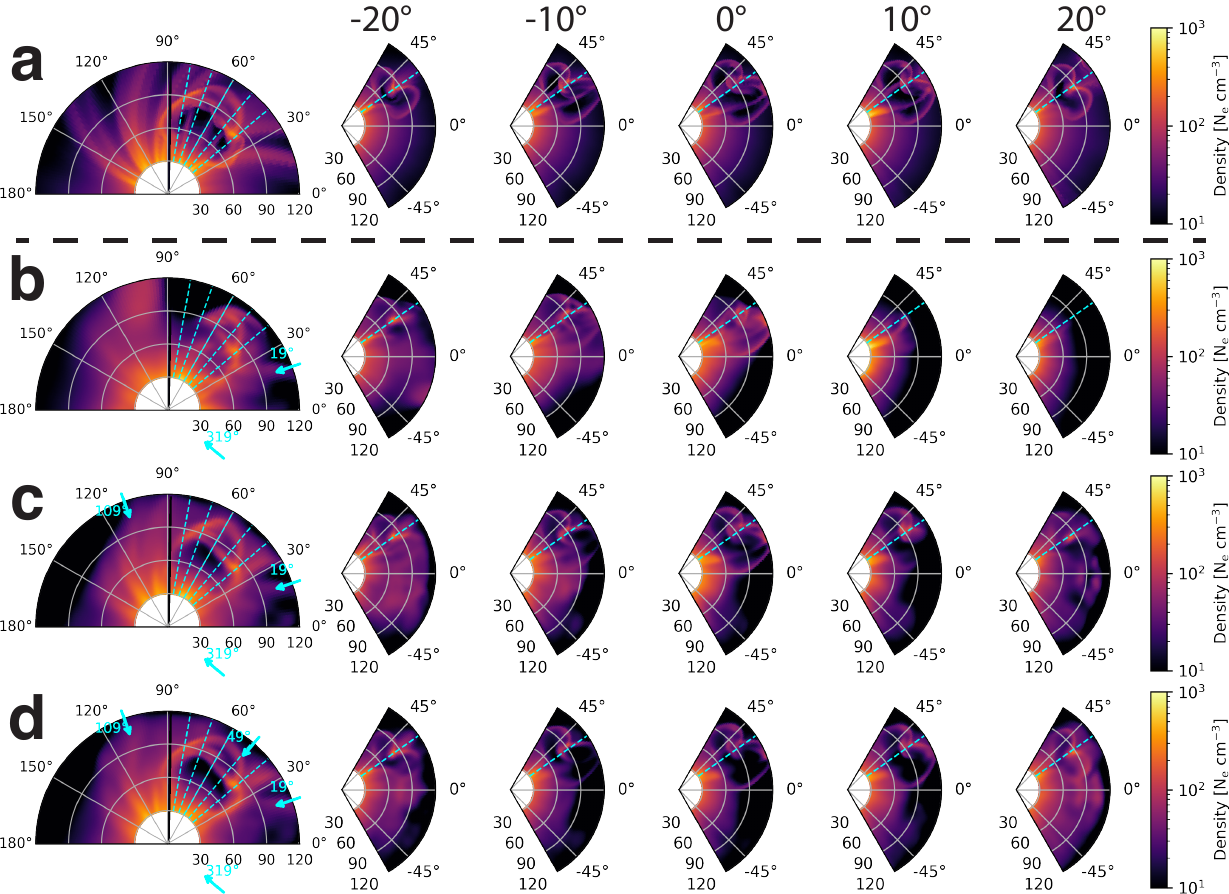}
    \caption{ Tomographic reconstruction of ``CME2'' from the CME Challenge.
(a) Ground-truth electron-density slices at $35^\circ$ latitude and $60^\circ$ longitude; the blue dashed lines mark the slice locations ($-20^\circ$, $-10^\circ$, $0^\circ$, $10^\circ$, $20^\circ$ relative to central longitude).
(b–d) Reconstructed slices obtained from two-, three-, and four-viewpoint configurations. Observer locations are indicated by blue arrows. Note that the latitude slices are not plane-parallel but correspond to conical surfaces at constant latitude.}
    \label{fig:cme2}
\end{figure}

The primary analysis focused on a single CME event, for which we varied the observer configuration to enable a direct comparison with the underlying ground truth. Here, we present additional reconstruction examples using different CME events from the CME Challenge dataset, comparing results obtained from two to four viewpoints, as available in the dataset. For these additional tests, the CME Challenge events are used strictly as independent cases: no ground-truth density information or prior event data were used for hyperparameter adjustment or model configuration. As in the previous evaluation, the reconstructions depend exclusively on the corresponding synthetic white-light images. 

Figure~\ref{fig:cme1} shows reconstructions of ``CME1'' from the CME Challenge, with the corresponding observer locations indicated by blue arrows. Reconstructions from two viewpoints recover the central part of the CME front (within approximately $\pm 10^\circ$), but the full extent is distorted due to interactions with the background solar wind. Adding a third viewpoint improves the reconstruction, revealing more of the extended shock front and providing clearer signatures across a wider range of latitudinal slices (up to $\pm 20^\circ$). Using all four available viewpoints yields the most complete reconstruction; however, even in this case the bright core and outer shock front remain more diffuse, reflecting the increased complexity of reconstructing a faint and spatially extended CME.

Figure~\ref{fig:cme2} shows reconstructions of ``CME2'' from the CME Challenge. As for ``CME1'', reconstruction quality improves consistently with the number of viewpoints. The correct identification and separation of background streamers remains challenging, especially with only two viewpoints. Nevertheless, adding a third viewpoint already leads to a clear improvement, which we attribute to the better localization of a prominent far-side streamer. ``CME2'' exhibits a more complex internal density structure than the CME analyzed in our main analysis, yet reconstructions from three and four viewpoints still reproduce the key morphological features, including the internal structure and overall shock-front geometry.

Taken together, these additional examples confirm the trends identified in the primary analysis. While the bright CME front can be reconstructed robustly, structures overlapping with the background solar wind are more difficult to disentangle and tend to appear smoother in the reconstructions. The successful application to multiple independent CME events demonstrates that the chosen training configuration and reconstruction strategy generalize beyond the specific case used for method development, supporting the broader applicability of the proposed framework.

\section{Ablation Study of Physics-Based Loss Weighting}
\label{sec:physics_ablation}

\begin{figure}
    \centering
    \includegraphics[width=0.7\linewidth]{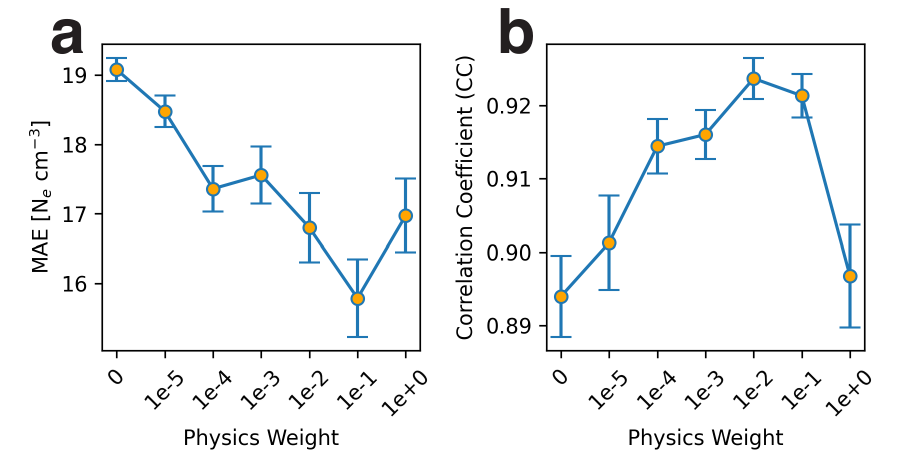}
    \caption{
Ablation study of the physics-informed loss weighting. 
(a) Mean absolute error (MAE) of the reconstructed electron density relative to the ground truth as a function of the physics loss weight. 
(b) Cross-correlation coefficient (CC) between reconstructed and ground-truth densities. 
Increasing the physics weight improves reconstruction quality up to an optimal range around $10^{-2}$--$10^{-1}$. 
For excessively large weights ($\lambda = 1$), the solution quality degrades as the optimization becomes dominated by the physical regularization terms at the expense of fitting the observations.
}
    \label{fig:physics_ablation}
\end{figure}

To further quantify the role of the physics-informed constraints, we perform a systematic ablation study of the corresponding loss weights. In contrast to the qualitative comparison presented in Sect.~\ref{sec:physics_constraints}, here we explicitly vary the weighting factors associated with the physical regularization terms. 
Specifically, we jointly vary the physics-informed loss components
\[
\lambda_{\rm continuity} = \lambda_{\rm radial} = \lambda_{\rm velocity}
\in \{0, 10^{-5}, 10^{-4}, 10^{-3}, 10^{-2}, 10^{-1}, 1\},
\]
while keeping the image-based loss terms fixed ($\lambda_{\rm continuity} = \lambda_{\rm ratio}=1$). For each configuration, we reconstruct the CME event using the same observational setup with three observers, analogously to Sect. \ref{sec:noise}, and compute the reconstruction accuracy relative to the ground-truth plasma distribution.

Figure~\ref{fig:physics_ablation} summarizes the resulting performance in terms of the mean absolute error (MAE) and cross-correlation coefficient (CC). The results demonstrate a systematic improvement in reconstruction quality as the physics-based regularization is increased from zero. Moderate weighting factors in the range $10^{-2}$ to $10^{-1}$ yield the best performance, indicating an optimal balance between observational fidelity and physical consistency.

For vanishing weights ($\lambda = 0$), the solution reduces to a pure NeRF formulation, which is more susceptible to projection artifacts and temporal inconsistencies. Conversely, when the physics weights become dominant ($\lambda = 1$), the reconstruction quality deteriorates substantially. In this regime, the optimization is overly constrained by the physical regularization terms, preventing the model from adequately fitting the observational data and leading to degraded solutions.

These results confirm that the physics-informed constraints act as effective regularizers, improving stability and accuracy when appropriately balanced with the observational loss terms. Importantly, we observe a well-defined transition regime: reconstruction quality improves monotonically up to intermediate weights, followed by a clear degradation once the physics terms become dominant. This sharp cutoff indicates that the optimization exhibits a distinct stability boundary rather than a gradual drift, which is favorable for selecting suitable weighting factors in practice. The existence of this regime suggests that the optimal balance between data fidelity and physical consistency can be robustly identified.



\bibliography{cite}{}
\bibliographystyle{aasjournal}



\end{document}